\documentclass[sigconf,pbalance]{acmart}

\usepackage{amsmath,amsfonts}
\usepackage{algorithmic}
\usepackage{graphicx}
\usepackage{textcomp}
\usepackage{xcolor}
\usepackage{comment}
\usepackage{tikz}
\usepackage{subcaption}
\usepackage{multirow} 
\usepackage{enumitem}

\newcommand{\mytexttilde}{\raisebox{0.5ex}{\texttildelow}}

\newcommand*\circled[1]{\tikz[baseline=(char.base)]{\node[circle, fill=black, inner sep=0.1ex, text=white] (char) {\scalebox{0.8}{#1}};}}

\newcommand*\circledsmall[1]{\tikz[baseline=(char.base)]{\node[circle, fill=black, inner sep=0.1ex, text=white] (char) {\scalebox{0.95}{#1}};}}

\newcommand*\filledcircledw[1]{\tikz[baseline=(char.base)]{\node[circle, fill=white, inner sep=0.1ex, text=black,draw=black] (char) {\scalebox{0.95}{#1}};}}

\usepackage{acmart-taps}
\aptLtoXcmd{
\long\def\circled#1{\xbox{aptbox}{\XMLaddatt{style}{border: 2px solid black;background-color: \#000000;color:\#FFFFFF;border-radius: 50\%; padding: 0.2em 0.5em;text-align:center;}{#1}}}}{}

\aptLtoXcmd{
\long\def\circledsmall#1{\xbox{aptbox}{\XMLaddatt{style}{border: 2px solid black;background-color: \#000000;color:\#FFFFFF;border-radius: 50\%; padding: 0.2em 0.5em;text-align:center;}{#1}}}}{}

\aptLtoXcmd{
\long\def\filledcircledw#1{\xbox{aptbox}{\XMLaddatt{style}{width: 15px; height: 15px; border: 1.2px solid black; border-radius: 50\%; padding: 0.2em 0.5em;text-align:center;
}{#1}}}}{}

\newcommand{\sysname}{\protect{\sc EcoCenter~}} 
\newcommand{\sysnamens}{{\sc EcoCenter}}
\newcommand{\sysnamesec}{{EcoCenter}}

\copyrightyear{2026}
\acmYear{2026}
\setcopyright{cc}
\setcctype{by}
\acmConference[ICS '26]{2026 International Conference on Supercomputing}{July 06--09, 2026}{Belfast, United Kingdom}
\acmBooktitle{2026 International Conference on Supercomputing (ICS '26), July 06--09, 2026, Belfast, United Kingdom}
\acmDOI{10.1145/3797905.3807852}
\acmISBN{979-8-4007-2522-7/2026/07}

\begin{document}

\title{Coordinating GPU Data Centers and Power Grid Regulation Service for Exogenous Carbon Benefits}

\author{Ali Jahanshahi}
\affiliation{%
\department{Department of Computer Science and Engineering}
   \institution{University of California, Riverside}
   \city{Riverside}
   \state{CA}
   \country{USA}
}
\email{ajaha004@ucr.edu}

\author{Sara Rashidi Golrouye}
\affiliation{%
\department{Department of Computer Science and Engineering}
   \institution{University of California, Riverside}
   \city{Riverside}
   \state{CA}
   \country{USA}
   }
\email{srash034@ucr.edu}

\author{Osten Anderson}
\affiliation{%
  \department{Department of Electrical and Computer Engineering}
   \institution{University of California, Riverside}
   \city{Riverside}
   \state{CA}
   \country{USA}
   }
\email{oande001@ucr.edu}

\author{Nanpeng Yu}
\affiliation{%
  \department{Department of Electrical and Computer Engineering}
   \institution{University of California, Riverside}
   \city{Riverside}
   \state{CA}
   \country{USA}
   }
\email{nyu@ece.ucr.edu}

\author{Daniel Wong}
\affiliation{%
  \department{Department of Electrical and Computer Engineering}
   \institution{University of California, Riverside}
   \city{Riverside}
   \state{CA}
   \country{USA}
   }
\email{danwong@ucr.edu}

\begin{abstract}

The rapid growth of AI/ML data centers has led to higher energy consumption and carbon emissions. The shift to renewable energy and growing data center energy demands can destabilize the power grid. 
Power grids rely on \textit{frequency regulation reserves}, typically fossil-fueled power plants, to stabilize and balance the supply and demand of electricity. This paper sheds light on the hidden carbon emissions of frequency regulation service. 
Our work explores how modern GPU data centers can coordinate with power grids to reduce the need for fossil-fueled frequency regulation reserves. We first introduce a novel metric, \textit{Exogenous Carbon}, to quantify grid-side carbon emission reductions resulting from data center participation in regulation service.
We additionally introduce \textit{EcoCenter}, a framework to maximize the amount of frequency regulation provision that GPU data centers can provide, and thus, reduce the amount of frequency regulation reserves necessary. 
We demonstrate that data center participation in frequency regulation can result in Exogenous carbon savings that can outweigh operational carbon emissions.  
\end{abstract}



\maketitle

\section{Introduction}\label{sec:intorduction}
Data centers are essential infrastructures for supporting cloud services and modern ML/AI workloads. Data centers are now one of the leading electricity consumers worldwide. In the United States, data centers accounted for \mytexttilde 4\% of electricity consumption in 2024, and as high as 9.1\% by 2030~\cite{aljbour2024powering}. For certain states, such as Virginia, data centers account for 25\% of the state's total electricity consumption, with this projected to grow to 46\% by 2030~\cite{aljbour2024powering}.
This rising electricity demand increases carbon emissions. By 2030, global greenhouse gas emissions from data centers can be 40\% of total US greenhouse gas emission~\cite{morgan}. This growing environmental impact has led industry leaders to minimize carbon footprint of data centers. For instance, Google and Microsoft are committed to powering data centers with only carbon-free energy by 2030~\cite{google_report,ms_report}. 

Carbon footprint includes \textit{Operational Carbon} emissions, such as electricity consumption, and \textit{Embodied Carbon} emissions, such as supply chain activities. A major effort to reduce operational carbon is to shift towards renewable energy sources. Despite investments in renewable energy---an increase from 12\% to 31\% in carbon-free generation~\cite{renewable_data}---cloud providers continue to face a mismatch between their substantial data center demand and the available renewable energy supply.

However, both AI/ML data centers and renewable energy sources can cause instability to the power grid's frequency~\cite{jahanshahi2022powermorph,google_demand_response}. Power system operators need to dynamically balance energy generation and energy consumption to maintain grid stability through \textit{frequency regulation services}, where \textit{frequency regulation reserves} are expected to adjust their power levels based on 2-second granularity \textit{frequency regulation signals}. Since the intermittent nature of renewable energy sources cannot adjust output quickly enough to balance the grid, power grids still rely on fossil fuel-based power plants as a frequency regulation reserve. 

Both AI/ML data centers and \textit{renewable energy sources} increase the demand for frequency regulation reserves. Insufficient frequency regulation reserves can (1) impede the growth of future AI/ML data centers or (2) cause the power grid to intentionally reduce the amount of renewable energy available in the grid, leading to worse carbon emissions. \textbf{\textit{This paper aims to shed light on the hidden carbon emissions cost of power grid frequency regulation reserves, and how data centers providing frequency regulation service can help lower this.}} Because data centers account for a significant fraction of total grid energy consumption, their power flexibility holds great potential for reducing the demand for traditional frequency regulation reserves, thereby (1) supporting the growth of future AI/ML data centers and (2) enabling greater renewable energy penetration. Towards this goal, this paper makes the following main contributions:

\textbf{Contribution 1:} We propose \textbf{\textit{Exogenous Carbon}}, a novel metric that quantifies the grid-side carbon emission reduction due to data center frequency regulation services. Data center regulation service can reduce the demand of traditional frequency regulation reserves, thus, lowering the carbon emissions of the power grid. To the best of our knowledge, \textit{our work is the first to quantify the grid-side carbon emissions benefits of data center frequency regulation services.}

\textbf{Contribution 2:} We propose \textbf{\textit{EcoCenter}}, a framework to maximize  \textit{frequency regulation provision}, and thus, exogenous carbon savings of AI/ML GPU-based data centers. 
While prior works~\cite{jahanshahi2022powermorph,chen-energyqare:-2019} demonstrated data center regulation service by modulating CPU power, GPUs offer greater power flexibility (5x of CPUs), although their power management knobs present unique challenges.
Frequency regulation service is challenging due to fine-grain 2-second power modulation requirements and the limitations of GPU power management knobs~\cite{gpupower-cal}, such as limited accuracy and limited usable dynamic power range due to high active idle power. \sysname overcomes this by carefully coordinating power capping, core allocation, and multi-GPU coordination of co-located workloads.

\begin{figure}[!h]
    \centering
    \vspace{-1mm}
    \includegraphics[width=1\columnwidth,trim=0 5mm 0 0,clip]{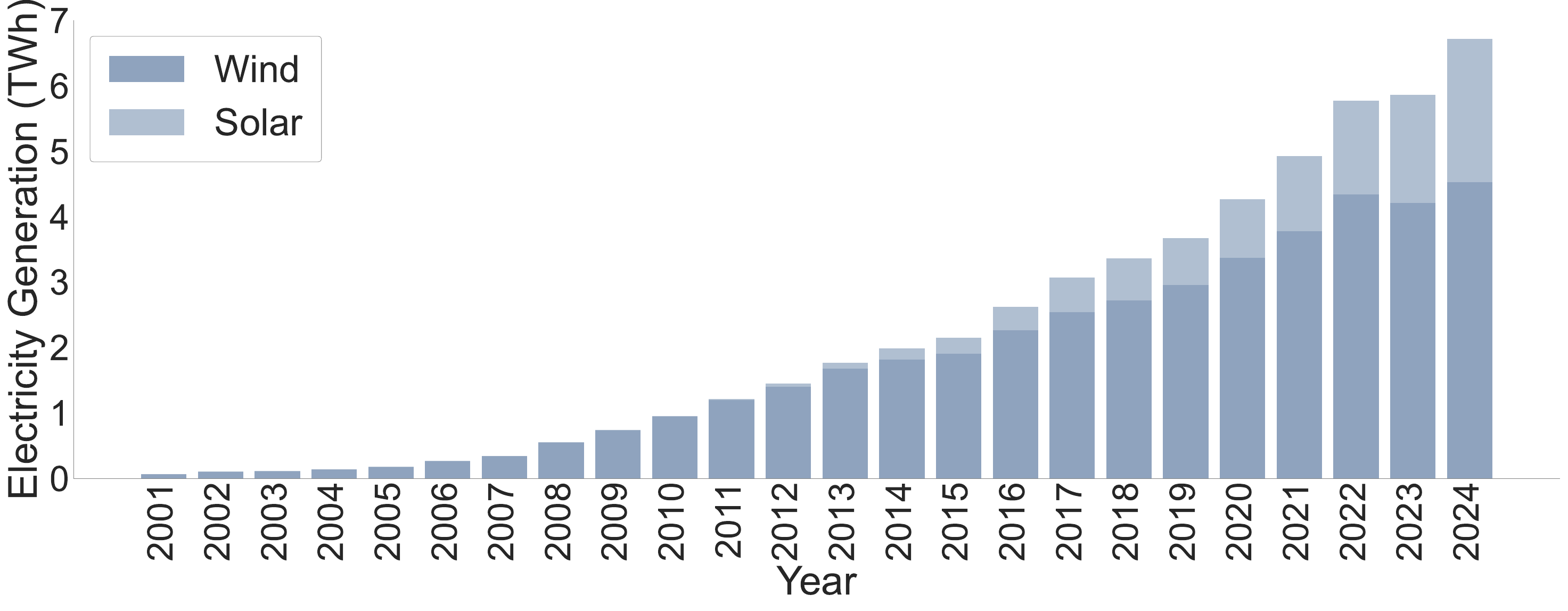}
    \vspace{-7mm}
    \caption{U.S. renewable solar and wind energy growth~\cite{renewable_data} can potentially destabilize electrical power grids~\cite{FreqRegNews}.} 
    \label{fig:renewable}
   \vspace{-4mm}
   \Description{Growth of Wind and Solar energy production in the United States.}
\end{figure}

\section{Background}\label{sec:background}
    \subsection{Carbon Emission Trends} \label{subsec:ml_cs_sustainability}

\textbf{Data Centers: } Recent hyperscale data centers consume upwards of 100 MW~\cite{cook2019clicking}. Globally, data center electricity is expected to increase by 160\% by 2030, with data centers consuming 3-4\% of worldwide electricity~\cite{GSprojection}.
Training and deploying machine learning models also contributes significant carbon emissions. For example, training the LLama2-70B language model produced 291.42 tons of CO\textsubscript{2} equivalent, despite state-of-the-art efficiency techniques~\cite{touvron2023llama}.

Recently, carbon-aware data center management techniques have been proposed to reduce the carbon footprint of ML workloads. 
For example, placing carbon-aware data centers in regions with greener energy sources~\cite{acun2023carbon-explorer}. Similarly, intelligent scheduling and coordination between data center operators and power grids can align flexible data center workloads with periods or regions of high renewable availability (i.e., low carbon intensity)~\cite{xing2023carbonresponder,dou2017carbon, luo2013temporal,limitations,li2023clover}. However, deeper integration between data centers and power grids is needed to reduce the carbon footprint of ML workloads.

    \textbf{Power Grids: } The need to reduce greenhouse gas emissions has made sustainable power grids a priority worldwide. In the United States, the integration of renewable energy sources, like solar and wind, increased by 54.5\% from 2018 to 2024, representing growth of more than half over these 6 years, as shown in Figure~\ref{fig:renewable}.
Due to the intermittent nature of renewable energy sources, this rapid integration poses significant challenges for grid stability to balance both electricity supply and demand~\cite{jahanshahi2022powermorph,google_demand_response}, requiring power grids to rely on demand response services to provide grid reliability. 

{\begin{figure}[h!]
    \centering
    \vspace{-1mm}
    \includegraphics[width=\linewidth,trim=0 4mm 0 0,clip]{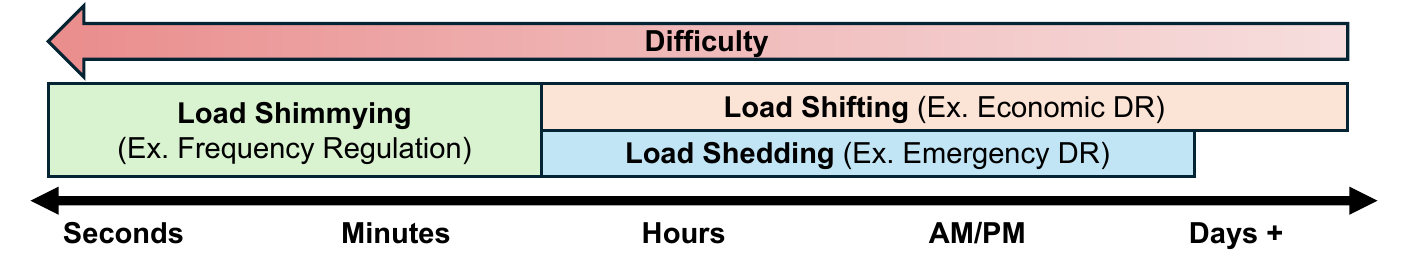}
    \vspace{-6mm}
    \caption{Various types of demand response (DR) services work across different timescales. Our work targets frequency regulation service, the most challenging DR service~\cite{LBNL-DR}. }
    \label{fig:DR-type}
    \vspace{-5mm}
    \Description{Illustration showing the time-scale of properties of different types of demand response services, such as load shimmying, load shifting, and load shedding.}
\end{figure}

\subsection{Demand Response Services}
Electricity market services provided by loads, such as datacenters, are \textit{demand response} (DR) services which rely on electricity consumers to adjust their load by following a signal from the power utility. As shown in Figure~\ref{fig:DR-type}, DR services can be  classified into three categories: Load Shifting, Load Shedding, and Load Shimmying~\cite{LBNL-DR}.  

Many prior works have explored data center participation in emergency demand response, a type of Load Shedding service operating over hours, which require data centers to reduce load during power grid emergencies~\cite{zhang2019data,google_demand_response,zhang2022hpc,xing2023carbonresponder,7039172}. Data centers shifting workloads to run during periods of greater renewable energy generation or cheaper electricity price are examples of Load Shifting service~\cite{Li2012-ISwitch,goiri2011greenslot,Li2011-SolarCore,Li2013-Chameleon,liu2011,dou2017carbon, luo2013temporal,lin2023adapting,CarbonScaler,10.1145/3698038.3698542, limitations}. 

The most challenging DR service is Load Shimmying, which requires participants to vary their load in minute or even second granularity. Examples include Load Following services that operate at 5-minute granularity, or advanced fast frequency regulation services, that operate at 2-second granularity (target of this paper).

    \begin{figure}[!t]
    \centering
    \includegraphics[width=\linewidth,trim={0 3mm 0 0},clip]{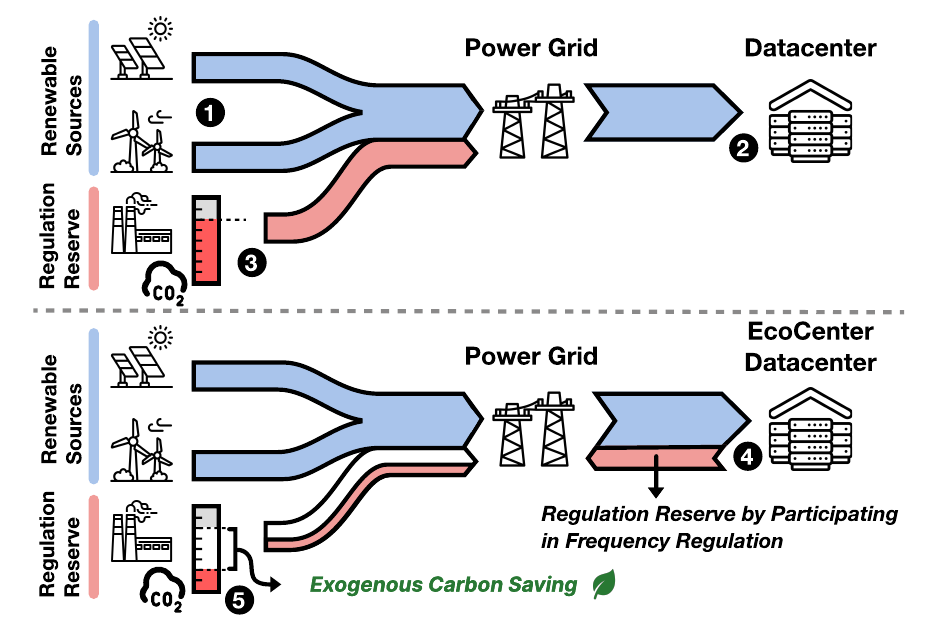}
    \vspace{-7mm}
    \caption{Power grids must balance \protect\circledsmall{1} the power generation of renewable resources and \protect\circledsmall{2} end-user power consumption by relying on \protect\circledsmall{3} fossil-fuel power plants for regulation reserves. This work explores \protect\circledsmall{4} how data centers can help grid balancing to reduce reliance on traditional regulation reserves and quantify the grid-side carbon footprint benefits \protect\circledsmall{5}.}
    \label{fig:gridDC}
    \label{fig:gridDC-fig}
    \vspace{-7mm}
    \Description{Illustrative figure showing the flow of carbon emissions sources and how power generator, data centers, and regulation reserves interact.}
\end{figure}

\subsection{Frequency Regulation Services}
\label{subsec:background-regulation-service}
To maintain electrical grid stability, operating frequency must be kept between 58.98Hz and 60.02Hz in the United States. As illustrated in Figure~\ref{fig:gridDC}, the power grid must maintain a balance between renewable energy generation \circledsmall{1} and power consumption \circledsmall{2}. However, the intermittent nature of renewable energy sources, like wind and solar, causes significant variation in power generation \circledsmall{1}. These sources cannot adjust output quickly enough to balance consumption. Traditionally, grid balancing is achieved by adjusting \textit{regulation reserve} generator output, typically by fossil fuel-based power plants, to match electricity consumption \circledsmall{3}. Specifically, these \textbf{\textit{regulation reserves can contribute a significant amount of carbon emissions that is not accounted for in traditional carbon intensity metrics of power generation sources}}~\cite{caiso-ghg,caiso-ghg2}. In Section~\ref{sec:exocarbon}, we will discuss this limitation further. 


\textbf{Frequency Regulation Markets:}
Power system operators offer \textit{frequency regulation services} in \textit{day-ahead} or \textit{real-time} markets. There are several Independent System Operators (ISO) that coordinates the regional movement of electricity in the USA. For example, PJM covers the eastern US and California ISO (CAISO) covers California. We utilize the CAISO real-time market, which allows bids every hour at 60-minute granularity. Resources, such as data centers, submit into the market every hour their estimated energy consumption baseline (P\textsubscript{avg}) and frequency regulation service provision capability (R), which represents the amount of power capacity (in MW) that the data center can commit for regulation.

\textbf{Regulation Signal:}
Regulation service resources, such as data centers, modulate their power consumption to follow a \textit{frequency regulation signal} (r(t)), which ranges from $[-1,1]$ and is broadcast every 2 seconds by the ISO. ISOs ensure that the difference between two consecutive values of r(t) does not exceed 0.5\% of R~\cite{pjm-pjm-2019}.

\textbf{Quantifying Quality of Frequency Regulation Service Provision:}
The revenue from frequency regulation service is dependent on the \textit{quality} of service provided. This quality is quantified by a \textit{performance score}~\cite{pjm-pjm-2019}, which is the average of delay, accuracy, and precision metrics. The performance score is calculated every 15 minutes by the ISO for each regulation resource~\cite{8318697}. ISOs certify a resource for regulation service provision after the resource achieves a performance score of 75\% or better on three consecutive tests~\cite{pjm-pjm-2019}. To maintain certification, resources must keep a performance score of at least 40\%, or be removed as a regulation resource.

\textbf{Reward Pricing and Electricity Cost:} 
Besides the electrical cost of the data center's electricity consumption, data centers also receive monetary rewards for participating in frequency regulation markets. 
By setting the energy consumption baseline (P\textsubscript{avg}) and the amount of frequency regulation provision (R), data centers maintain power consumption at $\mathrm{P_{avg} + r(t) \cdot R}$. The energy charge at time t equals the product of P\textsubscript{avg} and the locational marginal price of energy. The revenue from providing frequency regulation service equals the product of R, the quality of the provided regulation service, and the frequency regulation service price.

\textbf{Opportunities for Data Centers: }
With data centers projected to grow to 3-4\% of overall global power by the end of the decade~\cite{GSprojection} and accounting for up to 46\% of certain US state's electricity consumption~\cite{aljbour2024powering}, data centers can potentially provide a significant portion of frequency regulation capacity \circledsmall{4}, reducing the reliance on fossil fuel power plants \circledsmall{5}. Many prior works have explored the potential of data center participation in electricity markets~\cite{chen-energyqare:-2019,chen-data-2014,Chen2013,Chen2014,zhang2019data,jahanshahi2022powermorph,zhang2022hpc}.  
However, these prior studies were limited to analytical models or simulations, examined participation in slower demand response services or targeted data centers running best-effort workloads, which limits the applicability in real-world scenarios. Most relevant, PowerMorph~\cite{jahanshahi2022powermorph} explore how data centers running \textit{latency critical} workloads can participate in frequency regulation. However, PowerMorph only explore the economic benefits of frequency regulation and only modulates CPU power, an shrinking fraction of AI/ML data center power consumption~\cite{LLM-Power-Management}.

\setlength{\belowdisplayskip}{7pt} \setlength{\belowdisplayshortskip}{7pt}
\setlength{\abovedisplayskip}{7pt} \setlength{\abovedisplayshortskip}{7pt}

\section{Quantifying Grid-side Carbon Impact of Data Center Regulation Service}
\label{sec:exocarbon}

\textbf{Data Center Emissions Modeling: }
Data center carbon emission (C\textsubscript{DC}) is defined as the combination of \textit{operational} carbon (C\textsubscript{op}) and \textit{embodied} carbon (C\textsubscript{em}), $\mathrm{C_{DC} = C_{op} + C_{em} }$. 

\textit{Embodied carbon} accounts for the carbon footprint when building the data center and manufacturing the data center components. 
Following methodologies from prior work~\cite{li2023toward}, we derive \textit{embodied carbon}  from the aggregated manufacturing carbon footprint of the data center's hardware. 

\textit{Operational carbon} is the emissions produced from running a data center's day-to-day operations. This is typically based on the data center's energy consumption (E\textsubscript{DC}) and the carbon intensity (CI\textsubscript{gen}) of the power grid during data center operation, $\mathrm{C_{op} = E_{DC} \times CI_{gen}}$~\cite{acun2023carbon-explorer,li2023clover,lindberg2021guidereducingcarbonemissions,lacoste2019quantifyingcarbonemissionsmachine}.

\textit{Carbon intensity} metrics are provided by power grid operators and measures the \textit{average carbon emission} of the grid~\cite{electricitymaps,gorka2024carbonintensity}. 
The carbon intensity (CI\textsubscript{gen}) is computed as $\mathrm{CI_{gen} = \frac{C_{gen}}{E_{gen}}}$, where  C\textsubscript{gen} is the emissions from all energy generation sources and E\textsubscript{gen} is the amount of all energy generation. Carbon intensity is usually measured in units of gCO2eq/kWh and vary significantly based on the electricity mix and location~\cite{Sukprasert2024}. 

\textbf{Limitations of Data Center Emissions Modeling: } 
Note that the carbon intensity metric provided by power grids only account for \textit{power generation} and not \textit{regulation reserves}~\cite{caiso-ghg, caiso-ghg2}.
Therefore, the extent of existing carbon-aware data center interaction with power grids is to (1) reduce the data center's energy consumption (E\textsubscript{DC}), which reduces grid-side carbon emissions from power generation, or (2) schedule/relocate workloads to a different time/location with lower carbon intensity (CI\textsubscript{gen}).  
Clearly, existing modeling do not capture the emissions of grid-side regulation reserves, limiting the data center's ability to impact grid-side emissions.

\begin{figure}[!t]
    \centering
    \includegraphics[width=0.95\linewidth]{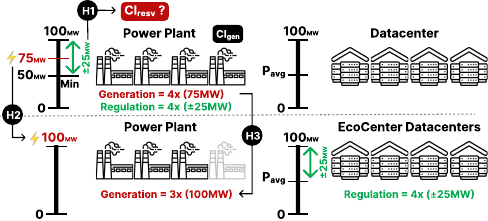}
    \vspace{-3mm}
    \caption{The ``hidden emissions'' of grid regulation reserves includes \protect\circled{H1} unaccounted emissions of \textit{performing} regulation service (grid only reports CI of generation), \protect\circled{H2} regulation reserves cause power plants to run at lower efficiency, and \protect\circled{H3} requires more grid resources to provide regulation capacity. Data center regulation service can alleviate these emissions.}
    \label{fig:gridDC-reserves}
    \vspace{-2mm}
    \Description{Hidden emissions in power plants and data centers. }
\end{figure}

\subsection{Accounting For Hidden Emissions of Regulation Reserves}
\label{sec:reserve-cost}
Figure~\ref{fig:gridDC-reserves} illustrates the interactions between grid frequency regulation reserves, data centers, and the ``hidden'' carbon emission factors that are currently unaccounted for in grid-reported carbon intensity metrics. This illustrative example has four 100MW data centers and four natural gas power plant with a minimum load.
Fossil-fuel power plants have a minimum stable operating load requirement in order to output power~\cite{power-min,power-min2,power-min3}.
In both scenarios, the grid is generating 300 MW of electricity with $\pm$100 MW of available regulation reserves. 

\noindent\circled{H1} \textbf{Hidden emissions 1: \textit{Providing} regulation service does not directly \textit{produce} emissions. }
Each power plant provides $\pm$25MW of regulation reserve. This reserve is only used when the grid requires balancing, thus, \textit{providing} reserves does not produce emissions, unless it \textit{performs} regulation.
This is the fundamental reason why \textit{\textbf{grid-reported carbon intensity cannot directly incorporate the emissions of regulation reserves}}.
Furthermore, performing regulation requires power plants to both \textit{increase} and \textit{decrease} output relative to nominal levels, which add challenges towards tabulating regulation emissions.

\noindent\circled{H2} \textbf{Hidden emissions 2: Power plants run at \textit{lower efficiency} to provide regulation service. }
To provide $\pm$25MW of regulation reserve, the power plants operate at 75MW to provide 25 MW of frequency regulation up and 25MW of frequency regulation down service. Power plants operate most efficiently at peak output~\cite{brooks2000ge}. Due to providing regulation, this power plant now operates at a less efficient output level of 75MW, which produce more emissions per MW of output.

\noindent\circled{H3} \textbf{Hidden emissions 3: Providing regulation service may require more power plants. } 
When data centers leverage workload flexibility to provide $\pm$25MW of regulation, it removes the grid-side carbon of performing regulation with fossil-fueled power plants and also allows power plants to operate more efficiently at 100MW output level. This also enable the dispatch of fewer power plants (3 instead of 4) in the grid to meet electricity demands. Therefore, data center participation in regulation service can have an out-sized carbon emissions benefit beyond the substituted regulation reserve.

\textbf{\textit{These fundamental limitations mean that existing operational carbon metrics inherently cannot capture grid-side regulation reserve emissions.}} This is because operational carbon only accounts for the carbon intensity of electricity generation (CIgen), completely ignoring the grid-side emissions from regulation reserves (the fossil-fuel power plants that provide the capacity to balance intermittent renewable energy, which is essential for grid stability). As a result, relying solely on operational carbon systematically misses this significant grid-side emission source. We need to fill this critical gap by explicitly quantifying these hidden grid-side emissions from regulation reserves, providing a more complete picture of data center carbon impact. 

\textbf{Capturing Regulation Reserve Emissions: } 
To capture these hidden emissions factors of regulation reserves, we extend operational carbon to include the impact of data centers on grid-side carbon emissions as following: 
{
\begin{equation}
\label{eq:operational_with_RS}
\mathrm{
    C_{op_{w/RS}} = E_{DC} \times CI_{gen} - \underbrace{(\mathrm{~R_{DC} \times MCE_{resv}})}_{\text{Exogenous carbon, } C_{exogenous}}
}
\end{equation}
}
, where R\textsubscript{DC} is the amount of regulation provision provided by the data center and MCE\textsubscript{resv} is the \textit{marginal carbon emission}~\cite{lindberg2021guidereducingcarbonemissions} of regulation reserves, which is the change in grid-side carbon emissions due to a change in regulation reserves in the grid. We call this new term \textit{exogenous carbon}\footnote{Exogenous is defined as ``relating to or developing from external factors'', hence, exogenous carbon relates to the various hidden carbon emission factors external to existing grid carbon intensity metrics. }, C\textsubscript{exogenous}, which 
collectively quantify the \textit{grid-side} carbon emission impact of data center participation in frequency regulation services. With this term, \textbf{\textit{data centers now have an additional knob for carbon-impact by maximizing the amount of regulation provision, R\textsubscript{DC}.}}

The next challenge is in how MCE\textsubscript{resv} can capture the aforementioned hidden emissions of regulation reserves. A simple model would be to only account for the carbon intensity of the power plant used for regulation, where $\mathrm{MCE_{resv} = CI_{resv}}$. However, this does not fully capture the power grid dynamics of ``hidden emissions'', resulting in under-estimation of exogenous carbon impact.

Since marginal carbon emission captures the \textit{change} in grid-side emissions due to a \textit{change} in regulation reserves, we require knowing the conditions of the grid with and without data center regulation service. Capturing all hidden emissions factor require a detailed power grid model that captures the emissions of \textit{performing} regulation service and the power plants dispatched (along with output levels) to provide electricity generation and regulation reserves. In electricity markets, these are achieved through a \textit{unit commitment} process.

\textbf{Obtaining MCE\textsubscript{resv} with Grid-side Unit Commitment Modeling: } The unit commitment problem in electricity markets is the process of determining which power generation units should be turned on/off to meet expected electricity demand, at the lowest cost, while satisfying technical and operational constraints. It involves scheduling generators over a 24 hour period to ensure a reliable and cost-efficient supply of electricity. For example, CAISO performs unit commitment in a day-ahead market to schedule generation for the next day. Adding data center regulation reserves significantly impacts unit commitment when scheduling grid resources.  

We formulate and solve the unit commitment problem as laid out in Anderson~\cite{Anderson2025},  which models CAISO and the Western Interconnection grid, totaling 14GW. We assume that 770MW of frequency regulation must be held at all times, which is 5.5\% of grid capacity for regulation provision. From historical electricity pricing and demand traces (circa 2022), the model solves at hourly granularity the generation unit commitment requirements to meet electricity demands and regulation reserves. Similar high-fidelity power grid simulation model is used by the California Public Utility Commission and California Independent System Operators, demonstrating its utility in power systems planning.

Using this model, we can identify the ``hidden'' cost of regulation reserves by first running the model with  regulation reserves provided by traditional generation sources to obtain the total carbon emissions of the grid. Then we run the model with data center regulation service to obtain the total carbon emissions of the grid with datacenter regulation service. The difference of both values provides the ``hidden'' grid-side carbon emission impact of data center frequency regulation, which we use to obtain the grid-detailed marginal carbon emission, MCE\textsubscript{resv}, for use in Equation~\ref{eq:operational_with_RS}.

\section{\sysnamesec~ Framework} \label{sec:sysname}

We now introduce \sysnamens, a framework for GPU data centers to maximize the amount of regulation provision (R\textsubscript{DC}) and maximize benefits to grid-side exogenous carbon savings. Unlike prior works that minimizes \textbf{datacenter-side} power \textit{consumption} to save \textit{operational} carbon, our work aims to maximize \textbf{datacenter-side} power \textit{modulation} to save \textbf{grid-side} emissions of regulation reserves, which we quantify with exogenous carbon in Equation~\ref{eq:operational_with_RS}. 

\sysname focuses on GPU data centers due to GPUs now dominating power consumption in data centers~\cite{LLM-Power-Management}, consuming 5x more than CPU power and thus providing the greatest source of workload flexibility and potential for regulation reserve provision. GPU TDP has grown significantly with each new generation, going from 300W (P100)\textrightarrow 400W (V100)\textrightarrow 700W (H100)\textrightarrow 1000W (B200) over the last decade~\cite{A100,H100,B200,touvron2023llama}.

\textbf{GPU challenges: } Compared to prior CPU-based data center regulation service~\cite{jahanshahi2022powermorph,chen-energyqare:-2019}, GPUs power modulation present unique challenges compared to CPUs. 
(1) CPUs provide significant low-power states optimized for low utilization that enable \textgreater 80\% of dynamic power range to be accurately modulated. Thus, CPU-based policies do not directly translate to GPUs. (2) Without a focus on low utilization low-power states, GPUs are designed for peak performance rather than idle efficiency, consuming significant power regardless of utilization—maintaining high static power even at idle (~60W per GPU), limiting dynamic power range available for modulation to <50\%.
(3) GPU power management has limited low-power states, leading to limited accuracy~\cite{gpupower-cal}. This non-power-proportional design means CPU-based regulation policies~\cite{jahanshahi2022powermorph,chen-energyqare:-2019} cannot be directly applied to GPUs, and requires GPU-specific techniques to overcome the static power barrier.

\subsection{System Overview}

\sysname overcomes GPU's power modulation limitations by coordinating GPU DVFS, compute unit scaling, and multi-GPU load assignment as novel power modulation knobs.
Figure~\ref{fig:framework} illustrates a high-level overview of \sysnamens. In our demonstration, \sysname is designed for containerized (Docker) data center environments, but can be generalized to other deployment styles. 

\textbf{Data center-side:}  On each server, a \textit{Controller} is deployed that reshapes the power consumption of servers running GPU workloads to adhere to the \textit{regulation signal} issued by the grid, alongside the average power of the server for the next hour (Server~P') and regulation provision the server is providing (Server~R),  which are determined by the \textit{Optimizer}. The \textit{Optimizer} decides the amount of regulation provision (Server~R) by each server based on the predicted average load (P\textsubscript{avg}) of the server and its variance (P\textsubscript{var}). We predict average load for the next hour based on historical load patterns~\cite{6881647,cetinski2015ame,di2012host,6274197,liu2017adaptive,8102182,8301555}.

\textbf{Power grid-side:} The grid provides the \textit{electricity cost} and \textit{regulation reward} to the data center and asks the data center to modulate its power based on the \textit{regulation signal}. At the end of each hour, the power grid credits the data centers with a monetary reward based on the \textit{performance score}, which is taken into account in electricity cost savings on the data center. 
We model the exogenous carbon savings (as discussed previously in Section~\ref{sec:exocarbon}) using the aggregate average power of the data center (Data~Center~P') and the aggregate data center regulation provision (Data~Center~R).

\textbf{Maximizing regulation provision: }
In order to maximize regulation provision, \sysname must have sufficient workload flexibility to modulate without impacting workload QoS. Workloads are either latency-critical (LC) or best-effort (BE). LC workloads tend to offer less flexibility in modulating power due to strict QoS requirements. Therefore, \sysname co-locates LC and BE workloads to enable multi-GPU servers more dynamic power range that can be modulated, which enables greater amount of regulation provision. 
Frequency regulation signals can request servers to increase power consumption beyond what the server's LC workload is currently consuming; thus, co-locating LC and BE workloads enables GPU-based servers to modulate power in both directions.

\begin{figure}[!t]
    \centering
    \includegraphics[width=\columnwidth,trim=0 8mm 0 0,clip]{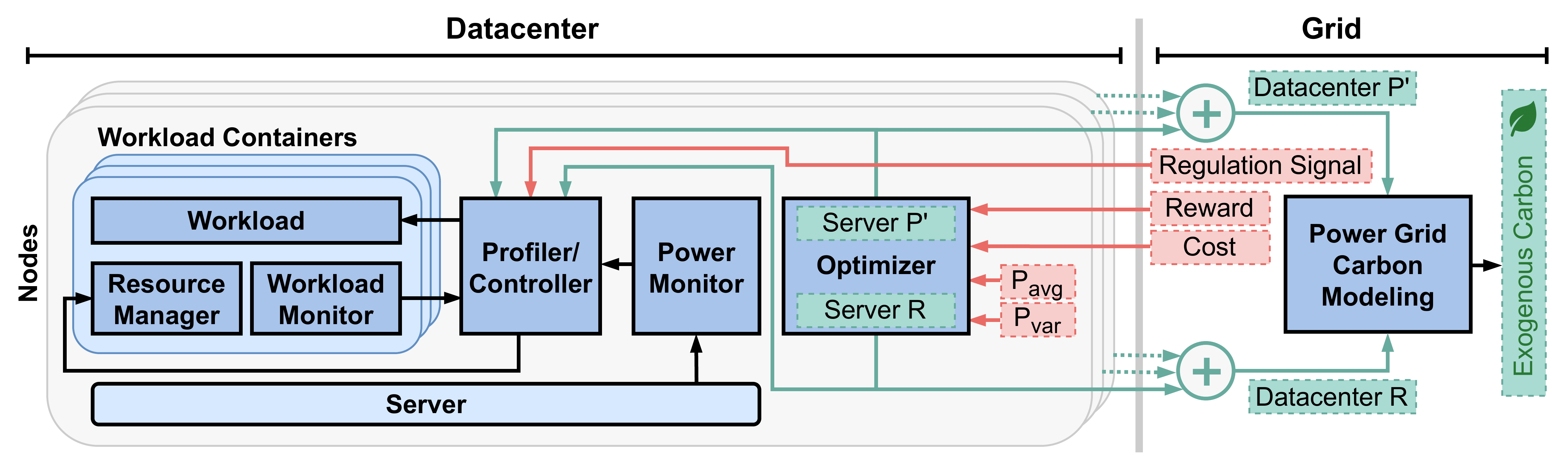}
    \vspace{-6mm}
    \caption{\sysname overview. Red boxes show the inputs to the framework and the green boxes show the output.}
    \label{fig:framework}
    \vspace{-5mm}
    \Description{EcoCenter overview figure. }
\end{figure} 
    \subsection{\sysnamesec~Workload and Power Monitor} \label{subsec:workload-monitoring} \label{subsec:power-monitor}

The \sysname \textit{Workload monitor}, a light-weight process within the workload's docker container, monitors the QoS of LC workloads and passes the performance metrics to the \textit{Controller} to make power reshaping decisions.
We used a multi-GPU inference server which load schedules the least amount of GPUs necessary for the current load~\cite{wattwiser23}. Thus, the workload monitor obtains the request inference times, resource requirements, and sends this to the \textit{controller}.

The \textit{Power monitor}, a dockerized service as shown in Figure~\ref{fig:framework}, uses GPU system management (SMI) API along with \texttt{Turbostat} tool, a CPU power measuring tool, to sample the power of the GPU and CPU, respectively. The power samples are published to its corresponding \textit{Controller} every second. The power samples are then used by the \textit{Controller} for power reshaping decisions.

    \subsection{\sysnamesec~Resource Manager} \label{subsec:resource-manager}

The \textit{Resource Manager} executes resource allocation decisions made by the \textit{Controller}. The \textit{Resource Manager} is a light-weight process within the workload's docker container and controls each GPU individually by sending different management commands to each. The \textit{Resource Manager} exposes two APIs to the \textit{Controller} for \textit{Power capping API} and \textit{Core allocation API}.

\begin{figure}[!t]
    \centering
    \includegraphics[width=0.85\columnwidth]{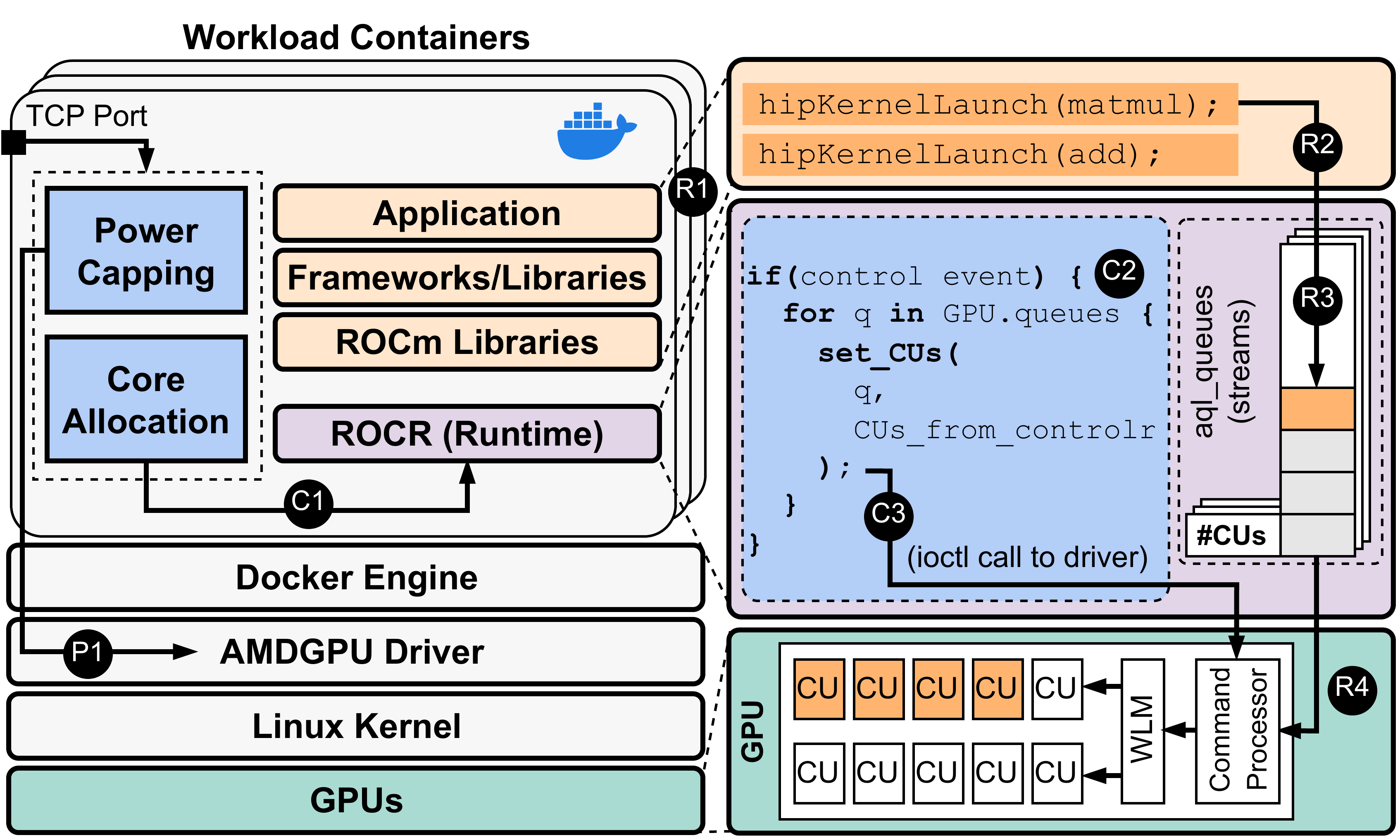}
    \vspace{-3mm}
    \caption{\sysname resource management. Dashed blue boxes show ROCm modifications to provide \textit{power capping} and \textit{core allocations} APIs to reshape server power. } 
    \label{fig:docker}
    \vspace{-6mm}
    \Description{ROCm-based runtime modifications to enable EcoCenter.}
\end{figure}

The \textit{Power capping} API, shown in Figure~\ref{fig:docker}, provides an interface to GPU system management interface (SMI)---nvidia-smi (for NVIDIA GPUs) and rocm-smi (for AMD GPUs)---to communicate with the GPU driver \circled{P1} and cap the power consumption of the GPUs with frequency scaling.

The \textit{Core allocation} API component is integrated into the framework to dynamically adjust GPU core allocations to each application. While our implementation uses AMD's CU Masking, EcoCenter's approach generalizes to NVIDIA GPUs through equivalent mechanisms: Green Context~\cite{nvidia_green_contexts} for dynamic resource partitioning and libsmctrl~\cite{libsmctrl} for fine-grained compute unit control. 
Alternatives, such as NVIDIA MPS and MIG provide coarse-grain resource partitioning at the process-level, which does not meet the kernel-level spatial partitioning requirements of our core allocation.
Figure~\ref{fig:docker} illustrates the integration of the \textit{Core Allocation} API into the ROCm runtime stack for GPU-accelerated workloads. 
Upon an application's \circled{R1} GPU kernel launch \circled{R2}, the kernel traverses the GPU runtime stack to be queued \circled{R3} into its designated stream residing in the ROCm runtime for execution on the GPU hardware \circled{R4}. Each queue contains a bitmask (CU Mask to designate core allocation of kernel), which defaults to all GPU cores. 

When the \textit{Controller} issues a core allocation command to the \textit{Core Allocation} API, a bitmask is generated by \textit{Resource Manager} and sent into ROCm runtime. To avoid costly Linux sysfs interfaces, we implement a shared memory interface between the \textit{Resource Manager} and ROCm runtime \circled{C1}. 
We modified the ROCm runtime to intercept kernel execution and manipulate their core allocation bitmask \circled{C2}. The bitmask is applied to all queued kernels through an IOCTL syscall \circled{C3}.

\begin{figure}[t!]
    \centering
    \includegraphics[width=\linewidth]{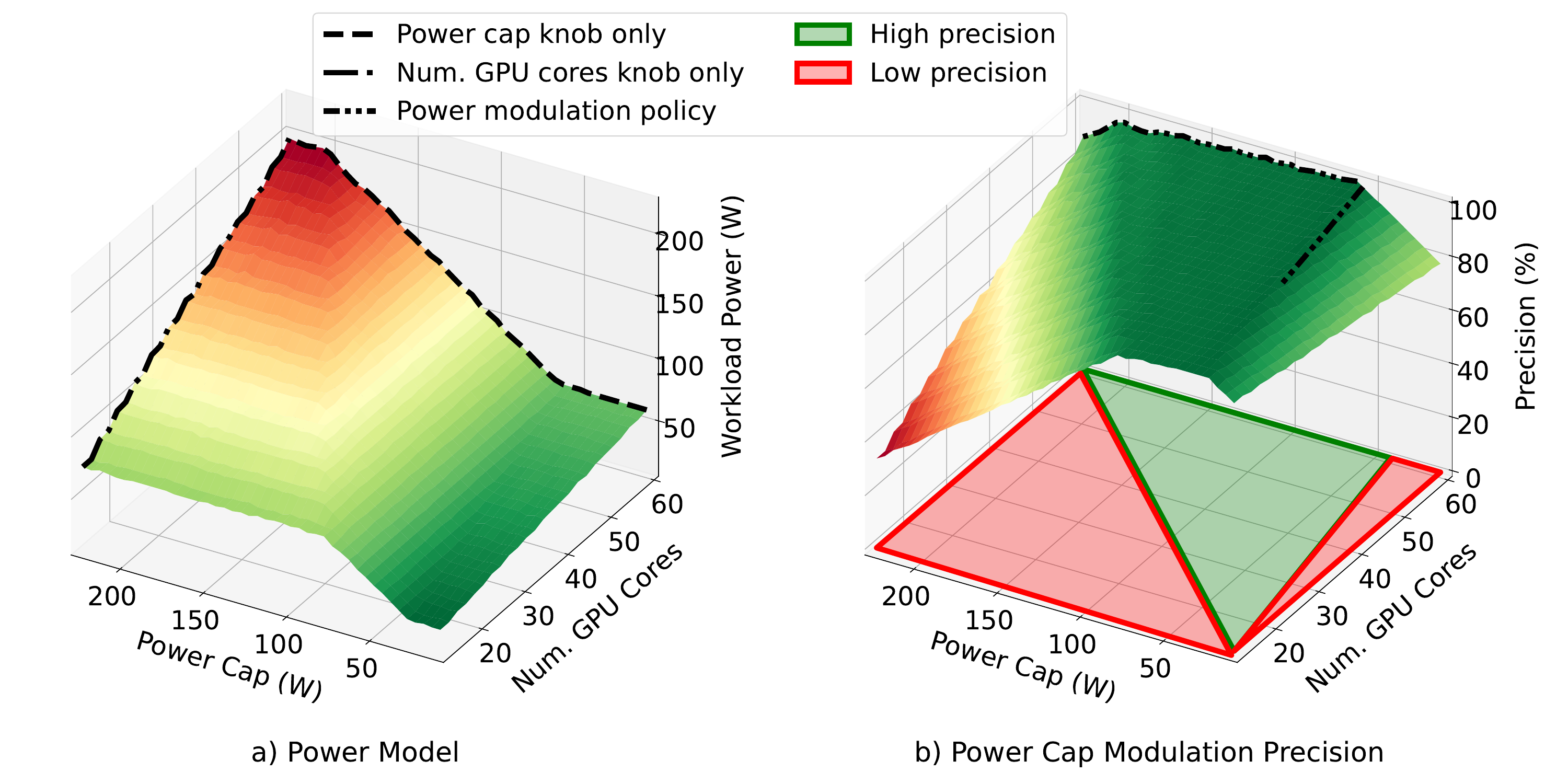}
    \caption{(a) GPT2 training power model (b) GPT2 model training power cap and number of GPU cores knobs precision (absolute difference of target power and consumed power).} 
    \label{fig:benchmark_gpt2-powermodel-tp}
    \vspace{-4mm}
    \Description{The power model and precision of power capping and resource scaling.}
\end{figure}
    \subsection{Power Modulation Knobs for GPUs} \label{subsubsec:power-reshaping}

To reshape server power, \sysname implements a power modulation policy to accurately follow the \textit{regulation signal}. To increase regulation provision, \sysname coordinates \textit{Power capping} and \textit{GPU core allocation} across all GPUs in a multi-GPU server. We first characterize a single GPU's power modulation knobs and investigate their impact on power modulation precision, workload throughput, and power efficiency.

\subsubsection{Power and Performance Model} \label{subsubsec:power-model}

\sysname constructs a power model for each BE workload, which we use to guide power modulation. 
Since we do not modulate the LC workload, we do not construct the LC power model and rely on real-time power monitoring. The \textit{Controller} adjusts power levels by utilizing the BE's power model to issue resource management commands to supplement the LC's real-time power usage, in line with the regulation signal.

We opportunistically construct the BE workload's power model online when there is GPU idleness. 
For a single idle GPU, we run the BE workload with varying power caps and CU masks, and extrapolate a multi-GPU power model. 
To build the power model, the Controller profiles power consumption data by varying the \textit{Power Cap} and \textit{GPU cores} while recording performance metrics for the workload. Figure~\ref{fig:benchmark_gpt2-powermodel-tp}(a) illustrates the power model for GPT2 training (BE) workload on a single GPU under various GPU cores and power capping constraints. Additionally, the Controller collects the workload's performance and throughput statistics under different resource constraints.

\subsubsection{Choosing the Power Modulation Policy}
In Figure~\ref{fig:benchmark_gpt2-powermodel-tp}(a), the dashed line shows the power modulation trend using only power capping. We observe that this knob is suitable for accurately modulating GPU power between \mytexttilde 60 -- 190 W, but not below \mytexttilde 60. Decreasing the power cap below this point does not significantly decrease GPU power. Note these power values are for AMD MI50 and would vary with other GPUs with similar power trends.

The dash-dotted line in Figure~\ref{fig:benchmark_gpt2-powermodel-tp}(a), shows the power trend of adjusting number of GPU cores with constant maximum power cap. We observe that this knob can modulate down to \mytexttilde 60W, but with low accuracy. 
From a performance view this knob impacts the performance of the workload more drastically, compared to \textit{power capping}, since workloads are more sensitive to resource count than operating frequency. We observe  core allocation knob degrades the workload throughput by up to \mytexttilde 80\% while power capping degrades up to \mytexttilde 40\%.  

\textbf{How to provide high quality power modulation?} 
We evaluate the precision (absolute difference of target power and actual consumed power) of using power capping to modulate the server power. Figure~\ref{fig:benchmark_gpt2-powermodel-tp}(b) illustrates the quality of the power capping knob. \textbf{\textit{To overcome the inaccuracies of GPU power knobs~\cite{gpupower-cal}, we use GPU core scaling to fine-tune the modulated power of power capping.}}
The red regions show power modulation precision \textless $90\%$ (low precision) and the green region with \textgreater $90\%$ (high precision). 
The power modulation policy aims to provide high quality power modulation, with $\geq90\%$ precision, within the green region. 

\textbf{How to navigate through high precision area to modulate power?} The difference in impact of \textit{core allocation} and \textit{power capping} knobs on workload performance/throughput is leveraged in power modulation policy implemented in \sysnamens.
Since power capping knob impacts on the workload's performance is half as much as GPU core counts, \sysname controller first modulates with power capping, then core allocation to fine-tune the power.

\textbf{Limitations of Single GPU Power Modulation Knobs: }
Following the above policies for individual GPUs, \textbf{\textit{each GPU still contains 60 W of untapped power range that can further contribute to regulation provision.}} 
Unlike CPUs, which include a plethora of low-power state knobs (DVFS, C-states, fine-grain power gating, preemption, etc.) to optimize non-peak utilization, a GPU's power knobs are limited to power capping and CU scaling and lack fine-grain power gating, which would otherwise provide a more dynamic range at lower power levels.
Thus, we require further mechanisms to increase regulation provision.

\subsection{Coordinating Multiple GPUs to Maximize Frequency Regulation Provision} \label{subsec:controller}

\begin{figure}[t!]
    \centering    
    \includegraphics[width=0.9\linewidth]{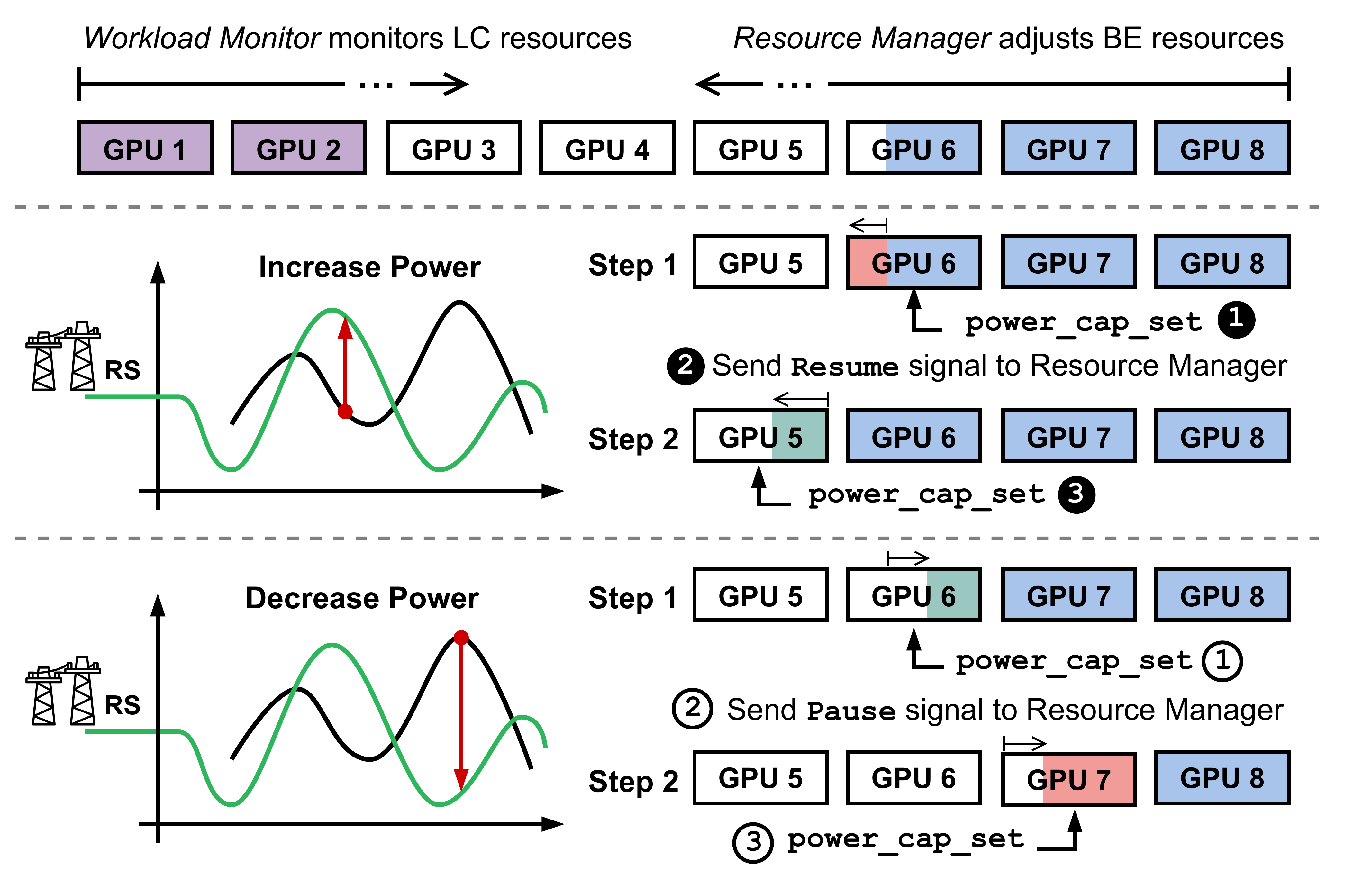}
    \vspace{-5mm}
    \caption{\sysname power reshaping policy with 8 GPUs. \textit{Workload Monitor} monitors the LC resources, while \textit{Resource Manager} (directly by \textit{Controller}) reshapes the BE workload. } 
    \label{fig:reshaping-policy}
    \vspace{-3mm}
    \Description{Illustrative example of how multiple GPUs are coordinated to improve the amount of regulation provision possible. }
\end{figure}

To maximize regulation provision with GPU's limited power modulation knob, the \sysname \textit{Controller} coordinates both LC and BE workloads, along with the GPU's power cap and core allocation knob. 
Although individually these knobs can achieve 2 second responsiveness, the challenge is to coordinate these across multiple GPUs to meet the 2-second responsiveness while maximizing regulation provision.

The \textit{Controller} reads the real-time power broadcasted by \textit{Power Monitor} and based on the regulation signal, it calculates the required target server power. 
Our assumption is that the LC workloads fluctuate unimpeded over time with varying number of GPUs to support the required load. Rather than co-locating other LC workloads to improve under-utilization, \sysname schedules BE workloads to fill in the under-utilization and modulates it for regulation service. Thus, to maximize the amount of regulation provision, \sysname carefully coordinate the GPUs by LC and BE groups.
Figure~\ref{fig:reshaping-policy} illustrates an overview of power reshaping policy. 

\textbf{LC and BE GPU allocation: } 
On the LC workload side, the \textit{Workload Monitor} is configured to allocate GPUs to the LC workload unobstructed (purple) from lower GPU index ($1$) to higher index GPUs ($8$) based on the current load of the server. 
For example, as the LC workloads fluctuate, it grows from GPU 1 to higher number GPUs. 
To avoid BE and LC interference, the BE workload's GPU resources are allocated starting from GPU with the highest index toward GPUs with lower index. For example, in Figure~\ref{fig:reshaping-policy}, BE workload's GPUs (shown in blue) are allocated from GPU $8$ towards lower indexed GPUs for modulation. 
The \textit{Workload monitor} track how many GPUs are required for the LC workload and dynamically free available GPUs to BE workloads. 
 
\textbf{Decreasing server power:} The controller decreases server power by issuing a \texttt{power\_cap\_set} command \filledcircledw{1} to \textit{Resource Manager} with a target power value lower than the current GPU power, until that GPU's minimum controllable power level. The issued target power value is calculated based on the profiled power model for the deployed BE workload.  The power can further be decreased by issuing a \texttt{pause} command \filledcircledw{2} to pause the GPU's workload, which can effectively remove the \mytexttilde 60W of untapped power range as previously described. Once paused, we can continue decreasing power by again calling \texttt{power\_cap\_set} command \filledcircledw{3} to the next GPU to further decrease server power.

\textbf{Increasing server power:} Conversely, Figure~\ref{fig:reshaping-policy}'s red sections show how server power is increased by adjusting GPU resources of the BE workload. The \textit{Controller} increases the server power by issuing a \texttt{power\_cap\_set}  command \circledsmall{1} to \textit{Resource Manager} with a target power value higher than the current GPU power. The power cap increments until reaching the GPU's maximum power. 

To further increase the power, the \textit{Controller} issues a \texttt{resume} command \circledsmall{2} to \textit{Resource Manager} to resume the BE workload processes running on the next GPU followed by a \texttt{power\_cap\_set} with the minimum power cap value for the GPU \circledsmall{3}. This results in an immediate \mytexttilde 60W jump in server power, so we adjust the other GPU's power accordingly to provide the necessary precision to meet server target power. 
 
\textbf{Resuming and Pausing BE Workloads: }
By coordinating multiple GPUs and resuming/pausing GPUs, we can tap into extra GPU power range (\mytexttilde 60W per GPU) that can be leveraged for regulation provision. 
To pause/resume individual GPUs, we run multiple single-GPU BE workloads, which is reflective of real clouds where most workloads are single GPU~\cite{weng2023beware}. When a training job on the GPU is paused, we stop issuing of new kernels to the GPU, which allows the GPU to go into an active idle state to achieve a lower near-zero power consumption. This is achieved through the ROCm runtime where we simply pause the issuing of kernels in the \texttt{aql\_queues} (Figure~\ref{fig:docker} \circled{R4}). 
After pausing, only the actively running kernel (with typical execution time in lower milliseconds) needs to drain.
When paused, memory state is retained in the GPU memory since the GPU is not completely powered off but in an active idle state. This way, the job can readily resume by reissuing kernels to the GPU.

    \subsection{\sysnamesec~Optimizer} \label{subsec:optimizer}

As mentioned in Section~\ref{subsec:background-regulation-service}, we utilize the CAISO real-time market. CAISO allows for asymmetric bidding and provides two different prices for regulation rewards: Up Regulation reward ($rew_{up}$) for decreasing power consumption, and Down Regulation reward ($rew_{down}$) for increasing power consumption.
We optimize the power consumption cost of a server participating in frequency regulation ($P_{f.r.}$) and the provided regulation provision by extending the optimization formulation proposed by \cite{jahanshahi2022powermorph} to provide asymmetric regulation provision ($R_{up}$ and $R_{down}$). Equation \eqref{eq:pick_pp} shows the formulation of the server with workload that consumes $P_{avg}$ power and the power variation (due to fluctuations in load) $P_{var}$.
{\footnotesize
\begin{equation}\label{eq:pick_pp}
    \begin{split}
        Maximize: \textbf{Saving} &= Cost_{P_{avg}} - Cost_{f.r.} \\ 
        such~that: Cost_{P_{avg}} &= \underline{P_{avg}}\times \underline{cost} \\ 
        Cost_{f.r.} = \textbf{P}_{\textbf{f.r.}} &\times \underline{cost_{~}} - Reward \times \underline{perf.~score}  \\
        Reward = R_{up} &\times \underline{rew_{up}} + R_{down} \times \underline{rew_{down}} \\ 
        Cost_{f.r.} &\leq Cost_{P_{avg}} \times \underline{threshold}\\
        0 \leq \textbf{R}_{\textbf{down}} &\leq \underline{P_{max}} - P_{f.r.} \\
        0 \leq \textbf{R}_{\textbf{up}} &\leq \textbf{P}_{\textbf{f.r.}} - (P_{avg} + \frac{P_{var}}{2}) \\
        \underline{P_{avg}} + \frac{{P_{var}}}{2} &< \textbf{P}_{\textbf{f.r.}} < \underline{P_{max}} \\
        \textbf{if}~symmetric~provision&:~\textbf{R}_{\textbf{up}}= \textbf{R}_{\textbf{down}} 
    \end{split}
\end{equation}  
}

In the optimization formulation of Eq.~\eqref{eq:pick_pp}, the \textbf{bold} parameters are outputs and \underline{underlined} parameters are inputs. The primary aim is to maximize savings by minimizing the difference between the average power cost ($Cost_{P_{avg}}$) and the frequency regulation cost ($Cost_{f.r.}$). The average power cost ($Cost_{P_{avg}}$) is determined by the product of the average power consumption ($P_{avg}$) and the unit cost of power ($cost$). On the other hand, the frequency regulation cost is influenced by the power allocated for frequency regulation ($P_{f.r.}$), the cost per unit power ($cost$), and a performance score factor (\textit{perf. score}). This cost is offset by rewards (\textit{Reward}) from regulation activities, which include the up or down regulation provision ($R_{up}$ or $R_{down}$). These rewards are multiplied by their respective reward values ($rew_{up}$, $rew_{down}$). 

Constraints ensure that the regulation cost doesn't exceed a specified \textit{threshold}, and that regulation provisions ($R_{up}$ and $R_{down}$) and power allocations remain within feasible bounds. 
The threshold parameter provides flexibility for data center operators to specify a threshold of savings to decide whether to participate in frequency regulation or withdraw from the service based on the state of the data center. 
The last 3 lines add constraints to ensure that the regulation provision upper range does not exceed the server's max power, the regulation provision lower range does not overflow into the LC workload's power range, and whether the ISO provides symmetric or asymmetric frequency regulation.
This formulation also can be utilized if symmetric provision is required, for markets such as PJM, where the regulation up ($R_{up}$) and down ($R_{down}$) capacities are enforced to be equal.

We utilize the Z3 SAT solver~\cite{de2008z3} to solve this optimization problem hourly to obtain the amount of regulation the data center will provide in the next hour. We observe that this optimization formulation can be solved by Z3 in the order of tens of seconds.

\section{Evaluation}\label{sec:evaluation}

\subsection{Evaluation Methodology}

\textbf{Experimental Setup and Benchmarks: }
We utilize a server with 8 AMD MI50 GPUs, a 16-core AMD EPYC 7302 processor, and 512GB of DRAM. The power consumption was monitored using rocm-smi~\cite{rocm-smi} and TurboStat. 

For our latency-sensitive workload, we run the Resnet152 model on a multi-GPU machine learning inference server~\cite{jahanshahi2022scaleserve} as a representative workload. We define the 95th percentile tail latency of the inference workload running independently by identifying the "knee" of the utilization-tail latency curve, similar to established methodologies~\cite{Chou2019-uDPM, Elfen}. For our best-effort workload, we use GPT-2~\cite{radford2019language} training. While we demonstrate with ML training, EcoCenter's power modulation knobs are hardware-based and application-agnostic, making our approach applicable to any batch-based BE workload. We measured the BE's throughput as training iterations per second (iter/sec). \textit{\textbf{Since \sysname does not modulate any latency-sensitive workloads, the goal here is to demonstrate that latency-sensitive workloads are untouched and should generalize to other latency-sensitive workloads.}}  
\textbf{\textit{As long as a BE workload can be modulated (our modulation knobs are application-agnostic and hardware-based), we can also generalize BE workloads.}} To test generalizability, we ran experiments with Albert inference (LC) and hipBLAS (BE) and observe similar trends, but omit detailed results for brevity. 

In our evaluation, the \textit{Baseline} scenario co-locates LC and BE workloads without regulation service, unless otherwise indicated. While the LC workload fluctuates, the BE workload is able to grow and fill up the under-utilized GPU resources in the server. The \textit{\sysname} scenario utilize the BE workloads for regulation service by modulating GPU power and multi-GPU coordination as discussed previously. For certain evaluations, we also evaluate a \textit{CPU-only regulation service} scenario similar to PowerMorph~\cite{jahanshahi2022powermorph},  which modulates a BE workload through CPU core isolation and DVFS. We also evaluate a \textit{PowerMorph-GPU} policy that naively applies PowerMorph's policy onto GPUs (which modulates a BE workload through CU isolation and does not support multi-GPU coordination), and a \textit{UPS-only regulation service} where regulation service is provided by the UPS battery backups.

\begin{table}[b!]
\vspace{-5mm}
\footnotesize
\centering
\caption{Workload utilization traces from the Facebook SWIM trace~\cite{swimprojectucb_2013} with combinations of load and variation.}
\label{table:trace}
\begin{tabular}{|l|c|c|c|c|}
\hline

\begin{tabular}[c]{@{}l@{}} \textbf{Trace} (\texttt{\textbf{Mean-Var}}) \end{tabular} &  \textbf{Mean(\%)}  & \textbf{Var.} & \textbf{Min (\%)} & \textbf{Max (\%)} \\ 
\hline
\hline
\texttt{low-low}        & 28.66 & 12.48 & 23  & 36\\
\texttt{low-med}        & 28.86 & 16.78 & 23  & 40 \\
\texttt{low-high}        & 29.33 & 22.62 & 24  & 42 \\
\hline
\texttt{med-low}        & 42.13 & 6.51  & 37  & 46\\
\texttt{med-med}        & 50.13 & 46.64 & 35  & 63 \\
\texttt{med-high}        & 51.66 & 177.55 & 36  & 73 \\

\hline
\texttt{high-low}        & 73.33 & 103.42 & 55  & 94\\
\texttt{high-med}        & 72.33 & 116.22 & 55  & 92 \\
\texttt{high-high}        & 70.73 & 144.99 & 52  & 91 \\

\hline
\end{tabular}
\end{table}

\textbf{Workload Utilization Traces: }
We evaluate under realistic varying workload utilization traces of Facebook SWIM project~\cite{swimprojectucb_2013} listed in Table~\ref{table:trace}. The traces are selected from 1-hour windows of the SWIM project trace to demonstrated a combination of \textit{load levels} where the mean load is low ($30\%$), medium ($50\%$), high ($70\%$), and the \textit{load variation} is low, medium and high variance.  
The LC workloads follow this utilization trace, while BE workloads either fill under-utilization (baseline) or track regulation signal.

\textbf{Regulation Signal Selection: }
We select three regulation signals from PJM, similar to prior work~\cite{jahanshahi2022powermorph}.  
Since we are participating in the hour-ahead regulation market, we picked one-hour slices. We chose Extreme (E) with a regulation signal that stays at the highest and lowest power points for extended periods; High Transition (HT) which features frequent min-to-max power change requests, and Noisy (N) to evaluate how accurately \sysname can track small frequent changes.

\textbf{Electricity Cost and Regulation Reward Selection: }
We analyzed the electricity rewards and regulation reward data from CAISO for the year 2022 and performed clustering to identify eight weeks that provide the best coverage of typical electricity market electricity cost and regulation reward across that year. We use this to evaluate Total Cost of Ownership (TCO) and electricity cost.

\textbf{Carbon modeling: }
We assume each GPU's carbon emission footprint is 30 kg CO\textsubscript{2}eq, with CPU, DRAM and disk having 18, 7, and 20 kg CO\textsubscript{2}eq, respectively~\cite{li2023toward}.  
To scale this to the data center-level, we scale the per-server carbon footprint by the number of servers that can fit within the data center's power capacity. We further amortize the embodied carbon over 5 years~\cite{li2023clover}. We assume a 100MW capacity data center in our evaluation. We also assume there exist sufficient UPS battery for 1 hour of backup with 100MWh of capacity with carbon footprint of 74 tons CO\textsubscript{2}eq per MWh~\cite{acun2023carbon-explorer}. 

For operational carbon (without regulation service), we utilize the CarbonExplorer~\cite{acun2023carbon-explorer} framework. 
We will evaluate both the \textit{simplified exogenous carbon model} and the \textit{detailed grid-based exogenous carbon model}. 
For the simplified exogenous carbon model's MCE\textsubscript{resv}, we evaluate two extreme cases where the regulation reserve is either a gas power plant or battery. Detailed exogenous carbon modeling was discussed in Section~\ref{sec:reserve-cost}.

\textbf{Data Center Scaling Methodology: }
Our evaluation uses measurements from a single 8-GPU server and scale up the frequency regulation impact to data center scale, similar to prior established methodology in evaluating data center frequency regulation service~\cite{jahanshahi2022powermorph}. We scale up to the number of servers in a 100MW data center from the per-server power. We scale three key metrics: \\
(1) Regulation provision: Each server independently calculates how much regulation provision (Server R) it can provide via Optimizer~\ref{subsec:optimizer} based on its workload. The data center aggregates this across servers to obtain a data center-scale regulation provision to bid for the next hour. 
The grid broadcasts a single regulation signal to data center.
Since each server bid their own regulation provision, the data center-scale regulation signal is proportionally scaled to each server's regulation provision and each server track the same data center-scale regulation signal pattern. Since each server independently tracks the regulation signal with consistent performance scores >80\% in Table~\ref{table:perf_score-ecocenter}, the aggregate data center performance remains >80\% and thus, we believe demonstration of a single server can accurately reflect data center-scale efficacy. \\
(2) Operational carbon scales by summing each server's energy consumption and applying the grid's carbon intensity.\\
(3) Exogenous Carbon scales with total regulation provision, calculated from the total Data Center R using the detailed grid-based exogenous carbon model. This approach combines server-level experimental validation with grid-level simulation.

\subsection{Evaluation Results}

\textbf{Performance Score: }
Table~\ref{table:perf_score-ecocenter} presents the average performance scores~\cite{pjm-pjm-2019} for providing regulation service. 
Recall, performance score is based on accuracy and precision of the server's power tracking the regulation signal and is a direct measurement of regulation tracking precision. 
The lowest performance score is observed with the Noisy regulation signal, which has the most intermediate (non-max or min) regulation signal values, which presents the most challenge for precise power modulation. 
High loads decrease performance scores due to less dynamic power range, making regulation signal tracking harder.
\sysname consistently achieves scores above 80\%, with an overall average of 91.20\%. 
We observed no regulation windows where performance score is below 75\%. 
This demonstrates that \sysname is able to accurately modulate power at 2-second granularity and provide high-quality regulation service. 

\begin{table}[t!]
    \centering
    \scriptsize
    \caption{\sysname regulation service performance score.}
    \label{table:perf_score-ecocenter}
    \begin{tabular}{|l|lll|l|}
    \hline
    \multicolumn{1}{|c|}{\multirow{2}{*}{\textbf{Load Trace}}} & \multicolumn{3}{c|}{\textbf{Regulation Signal}}                                    & \multirow{2}{*}{\textbf{Avg}} \\ \cline{2-4}
    \multicolumn{1}{|c|}{}                       & \multicolumn{1}{c|}{\textbf{\textit{Extreme (E)}}} & \multicolumn{1}{c|}{\textbf{\textit{Noisy (N)}}} & \multicolumn{1}{c|}{\textbf{\textit{High Trans. (HT)}}} &                          \\ \hline
    \hline
     \texttt{low-low} & \multicolumn{1}{l|}{96.72} & \multicolumn{1}{l|}{96.10} & 96.99 & 96.60\\ \hline
    
     \texttt{low-med} & \multicolumn{1}{l|}{96.94} & \multicolumn{1}{l|}{96.50} & 96.95 & 96.80 \\ \hline
    
     \texttt{low-high} & \multicolumn{1}{l|}{96.60} & \multicolumn{1}{l|}{96.06} & 96.84 & 96.5\\ \hline
    
    \texttt{med-low} & \multicolumn{1}{l|}{95.79} & \multicolumn{1}{l|}{95.56} & 95.50 & 95.62 \\ \hline
    
     \texttt{med-med} & \multicolumn{1}{l|}{93.02} & \multicolumn{1}{l|}{93.62} & 93.38 & 93.34 \\ \hline
    
     \texttt{med-high} & \multicolumn{1}{l|}{93.84} & \multicolumn{1}{l|}{93.64} & 94.69 & 94.05 \\ \hline
    
    \texttt{high-low} & \multicolumn{1}{l|}{85.36} & \multicolumn{1}{l|}{85.15} & 86.36 & 85.62\\ \hline
    
     \texttt{high-med} & \multicolumn{1}{l|}{84.10} & \multicolumn{1}{l|}{81.07} & 85.82 & 83.66\\ \hline
    
     \texttt{high-high} & \multicolumn{1}{l|}{84.94} & \multicolumn{1}{l|}{82.34} & 84.76 & 84.01 \\ \hline
    
    \end{tabular}
    \vspace{-2mm}
    \end{table}

\textbf{Workload Throughput: } Figure~\ref{fig:swim-throughput} illustrates the normalized throughput (y-axis) of both LC and BE workloads under different workload traces (x-axis) and regulation signals (series). Since LC workload's resources are preserved, the throughput is unaffected. For BE workloads, the workload is used to modulate the power to match the regulation signal. Since the workload is modulated, the BE workload naturally experiences reduced throughput compared to the baseline where the BE workload runs unmodulated. With higher workload load, the amount of regulation provision decreases, leading to higher throughput for BE workloads. Since we're modulating the workload around a regulation signal, one would expect the BE throughput to hover around 50\%. In general, we observe more throughput impact for Extreme and Noisy regulation signals, since the signal tends to be negative more often, requiring more throttling of BE workloads. In the remainder of the evaluation section, we conservatively evaluate on the Noisy regulation signal, as it is the most challenging, unless otherwise stated.

\begin{figure}[!h]
    \centering
    \includegraphics[width=1\linewidth,trim={0 30mm 0 0},clip]{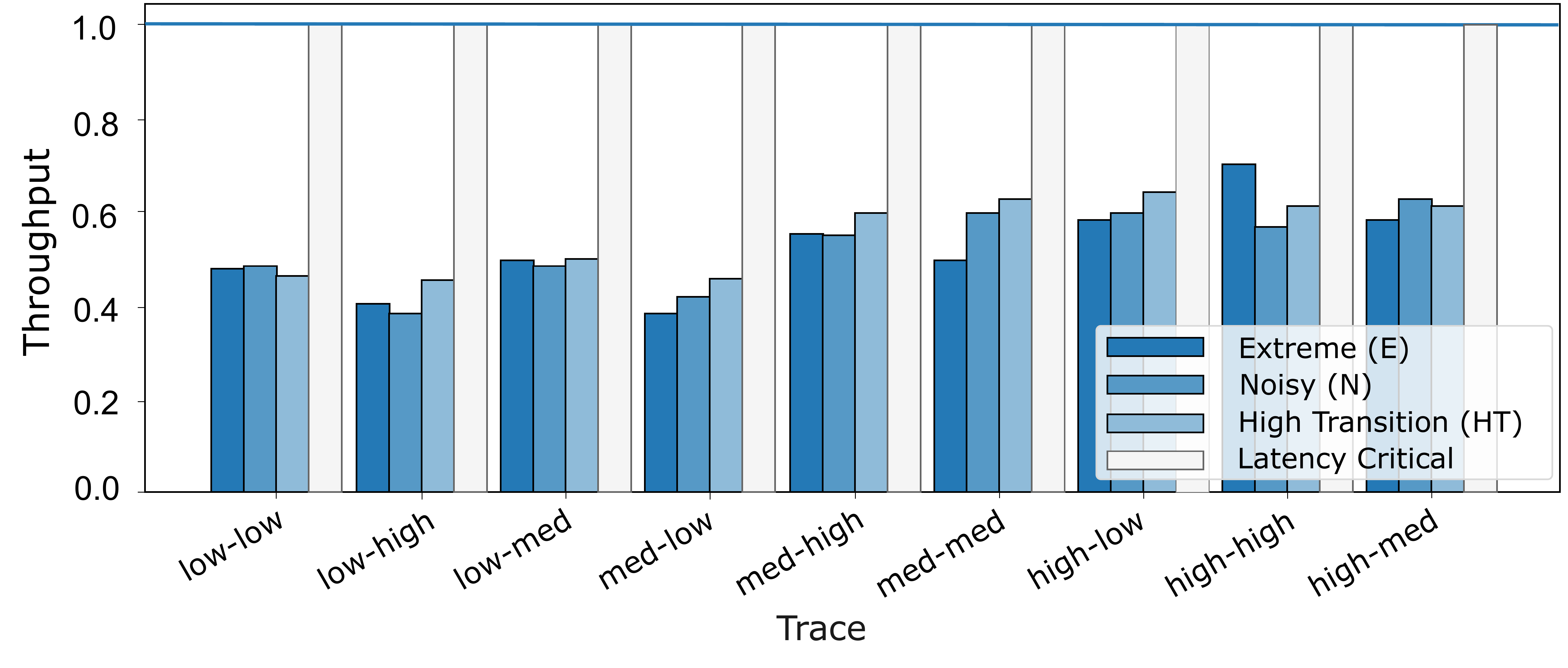}
    \vspace{-7mm}
    \caption{Normalized throughput for \sysname w.r.t the baseline with Facebook SWIM ~\cite{swimprojectucb_2013} traces. BE workloads are in blue and are sensitive to regulation traces. }
    \label{fig:swim-throughput}
    \vspace{-2mm}
    \Description{Normalized throughput with respect to the baseline.}
\end{figure}

\begin{figure*}[!t]
    \centering
    \includegraphics[width=0.85\linewidth]{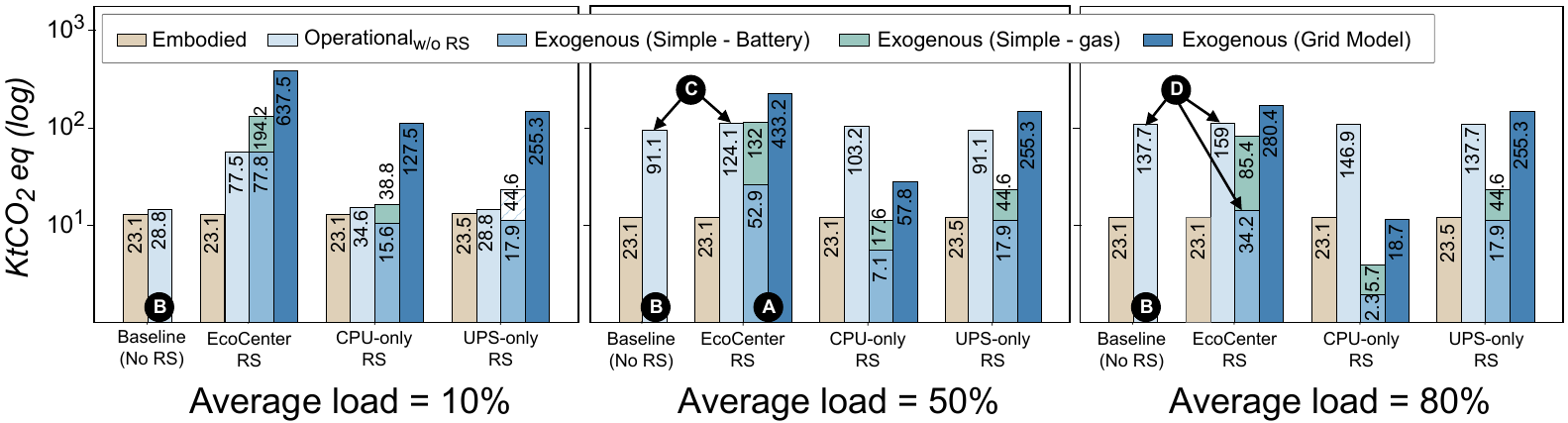}
    \vspace{-3mm}
    \caption{Comparison of Exogenous carbon savings with data center carbon emissions at different operational loads.}  
    \label{fig:exo-emb-op_carbon}
    \vspace{-2mm}
    \Description{Exogenous carbon savings at different operational loads.}
\end{figure*} 

 \begin{figure}[!t]
    \centering
    \includegraphics[width= 1\linewidth]{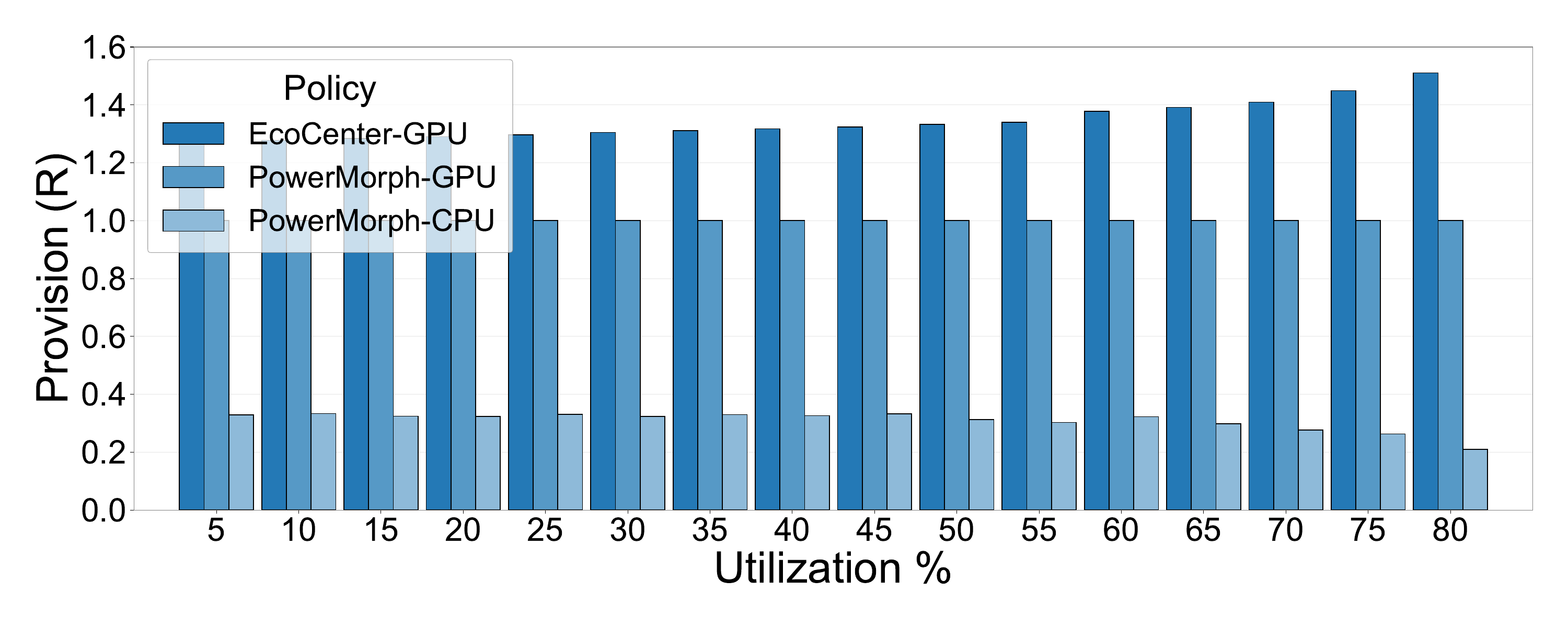}
    \vspace{-7mm}
    \caption{Amount of data center regulation provision normalized to PowerMorph-GPU. } 
    \label{fig:provision}
    \Description{Amount of data center regulation provision.}
    \vspace{-5mm}
    
\end{figure}

\textbf{Workload Latency: } Since the LC workload is not actively modulated and runs on a distinct set of GPUs compared to the BE workload, we do not see any noticeable differences in workload latency when running with and without regulation service. For brevity, we omit a latency figure.

\textbf{Regulation Provision: } Figure~\ref{fig:provision} shows the amount of regulation provision provided by \sysnamens, PowerMorph on CPUs, and PowerMorph on GPUs. All results are normalized to PowerMorph on GPUs. 
Because GPUs consume significantly more power than CPUs in data centers, \textit{\textbf{we observe that both GPU-based frequency regulation framework achieves \textgreater 3x more regulation provision than CPUs alone}}, providing significant power flexibility. 
As utilization increases, the amount of regulation provision shrinks. Going from 10\% to 50\%, and 50\% to 80\% utilization, the amount of regulation provision we observe (not illustrated in figure) decreased by 22\% and 56\%, respectively for PowerMorph-CPU, and 16\% and 35\%, respectively for PowerMorph-GPU, and 13\% and 26\% respectively for \sysnamens.  \sysname is able to coordinate across multiple GPUs to squeeze out regulation provision opportunities, allowing sustained regulation provision at higher loads compared to PowerMorph applied to GPU. \sysname provides \mytexttilde 30\% to 50\% more regulation provision compared to PowerMorph-GPU and 4x (up to 7x at high util.) compared to PowerMorph-CPU.

\textbf{Exogenous Carbon Saving: }
Figure~\ref{fig:exo-emb-op_carbon} shows the annualized carbon emission components for a data center with a 100MW capacity under various operational loads (10\%, 50\%, and 80\%). \sysname is compared to a \textit{baseline scenario where only the LC workload is running} so we can tease out the operational carbon due to the energy increase used for power modulation of the BE workload\footnote{If we assume the baseline scenario already runs co-located BE workloads to improve GPU utilization, then regulation service would observe reduction to both operational carbon \textit{and} grid-side exogenous carbon, with the trade-off being throughput (Figure~\ref{fig:swim-throughput}). We evaluate a Baseline of LC-only, which leads to a conservative estimate of \sysnamens's carbon emissions benefits.}. 
We also compare against a CPU-only \cite{jahanshahi2022powermorph} and a UPS-only regulation service (RS), where we assume 20\% of the UPS capacity is used for regulation service to keep battery capacity for backup. Embodied carbon is amortized over 5 years and remains consistent across all scenarios except for UPS-only RS due to the battery lifetime impact. Prior works observe that batteries deployed in regulation service experience a 28\% shorter lifespan~\cite{swierczynski_field_2014}. 

Similarly, to tease out the impact on grid-side regulation reserve emission benefits, we report operational carbon without regulation service, C\textsubscript{op}, and the exogenous carbon term, C\textsubscript{exogenous}, separately instead of the combined C\textsubscript{op\textsubscript{w/RS}} (Eq.~\ref{eq:operational_with_RS}).  
Operational carbon is estimated from the average datacenter power obtained from representative 10\%, 50\%, and 80\% load traces for LC workloads. Embodied carbon is independent of load. The load impacts the amount of regulation provision (Figure~\ref{fig:provision}) to modulate the BE workload based on regulation signals and thus, impacts Exogenous carbon savings.

As we discussed previously in Section~\ref{sec:reserve-cost}, when data centers provide regulation service, we see an out-sized impact to grid-side carbon emissions savings. We observe this where the grid model-based exogenous carbon benefits are often significantly higher (3x) than a simple exogenous carbon model (e.g. \circled{A} 433.2 kt CO\textsubscript{2}eq vs. 132 kt CO\textsubscript{2}eq for EcoCenter RS at 50\% load.), highlighting the need to consider high-fidelity modeling of grid-side dynamics. 

Operational carbon increases with higher loads (e.g. \circled{B} 28.8 kt CO\textsubscript{2}eq vs 91.1 kt CO\textsubscript{2}eq vs 137.7 kt CO\textsubscript{2}eq for Baseline as load increases). 
With EcoCenter and CPU-only RS, modulating co-located BE workloads also leads to operational carbon increase of running BE workloads compared to a baseline with LC-only workloads. For example, with 50\% load, the operational carbon \circled{C} increases from 91.1 to 124.1 kt CO\textsubscript{2}eq due to BE workloads for modulation. 
This increase is \textit{completely outweighed by the grid-side exogenous carbon savings} of \circled{A} 52.9-433.2 kt CO\textsubscript{2}eq. 
\textbf{\textit{In all scenarios, grid-side exogenous carbon savings completely outweighs the operational carbon increase for regulation service}}, matching or improving on the emissions of an LC-only baseline. 

Exogenous carbon savings decline as the load increases, reflecting reduced flexibility in providing regulation provisions (Figure~\ref{fig:provision}). Even in the worse-case with 80\% load, EcoCenter's grid-side exogenous carbon saving of 34.2 kt CO\textsubscript{2}eq \circled{D} still outweighs the operational carbon increase of 21.3 kt CO\textsubscript{2}eq, resulting in net carbon emission improvement of 12 kt CO\textsubscript{2}eq. In general, \textit{\textbf{the exogenous carbon benefits of CPU-only RS is very limited, and degrades quickly with load increase.}} While UPS-only RS outperforms CPU-only RS in exogenous carbon benefits, it still lags EcoCenter.

In certain scenarios (at 10\% load), grid-side exogenous carbon benefits \textit{completely outweigh} the datacenter’s carbon footprint, where the amount of grid-side exogenous carbon saved exceeds the data center’s operational and embodied carbon footprint. At low load, data centers offer a significant amount of regulation provision.

\begin{figure}[!t]
    \centering
    \includegraphics[width=1\linewidth,trim={7mm 6mm 11mm 15mm},clip]{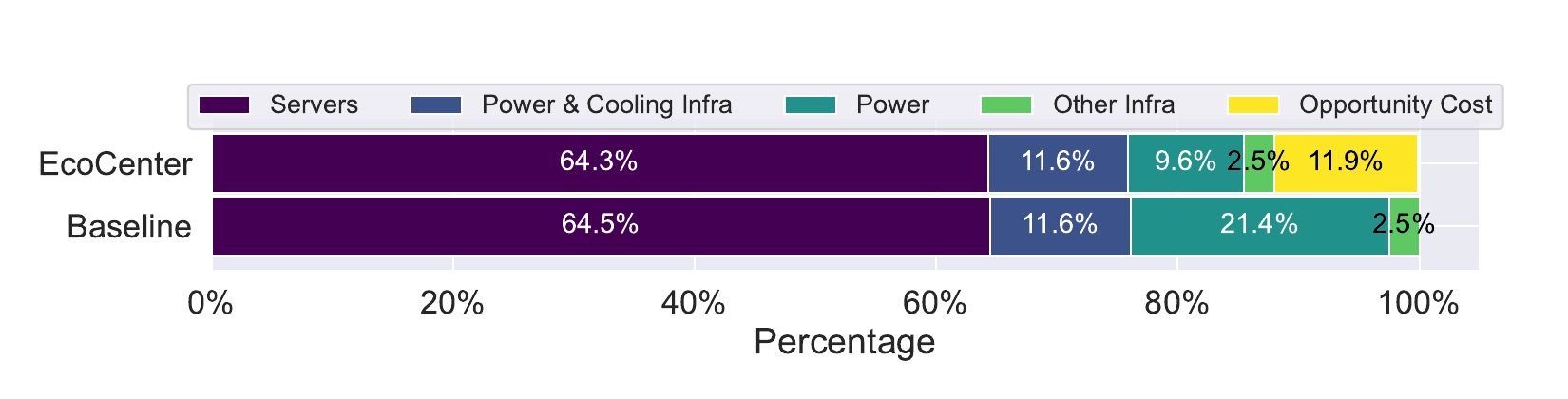}
    \vspace{-4mm}
    \caption{TCO breakdown showing CapEx (Servers, Infra), OpEx (Power), and Opportunity cost. }
    \label{fig:TCO-stacked}
    \Description{Total cost of ownership breakdown.}
\end{figure}

\textbf{Total Cost of Ownership: } To evaluate TCO, we follow the methodology of~\cite{hamilton,41606} and model a 100MW data center with facility construction cost of \$8 per W~\cite{41606} amortized over 15 years and PUE of 1.09~\cite{google_PUE}. We model server cost as state-of-the-art 8-GPU servers costing \$235K each requiring 7KW per server~\cite{exxact}. We evaluate a range of electricity cost from \$0.05 to \$0.30 per kWh. For reward pricing, through historical CAISO electricity and reward pricing, we find the average reward pricing for a certain electricity cost level, which we observe to average around \$7/MW~\cite{caiso-as}. 
Since regulation service modulates BE workloads, we capture this lost throughput as \textit{opportunity cost}. We model opportunity cost using the GPU hours lost and the hourly cost of a cloud GPU instance. 

Figure~\ref{fig:TCO-stacked} shows the cost breakdown of data centers with and without \sysnamens. In general, capital expenses (servers, power and cooling infrastructure, other infrastructure) remains the same. Operational expenses differ due to data centers receiving monetary rewards for participating in regulation service, hence the lower power cost component. However, this comes at the cost of opportunity cost when \sysname trade-off BE workload throughput to modulate power. This fundamental trade-off between opportunity cost and power cost (OpEx) drives the overall TCO picture.

Figure~\ref{fig:TCO-heatmap} shows the TCO of \sysname normalized to baseline. As electricity cost increases, it becomes a bigger component of TCO and thus can benefit from more rewards pricing, resulting in lower TCO. As the cost per GPU hour increases, we observe that opportunity cost begins to dominate and eventually outweighs the OpEx benefits and observe negative TCO impact. 
Therefore, to limit opportunity cost, frequency regulation is best done with lower cost per GPU hour resources while higher cost per GPU hour resources can still be dedicated towards latency-sensitive workloads, such as inference, without impacting its performance. 
Therefore, to limit opportunity costs, frequency regulation is best performed using resources with a lower cost per GPU hour, while resources with a higher cost per GPU hour can be dedicated to latency-sensitive workloads, such as inference, without impacting their performance.

\begin{figure}[!t]
    \centering
    \includegraphics[width=\linewidth,,trim={7mm 7mm 25mm 6mm},clip]{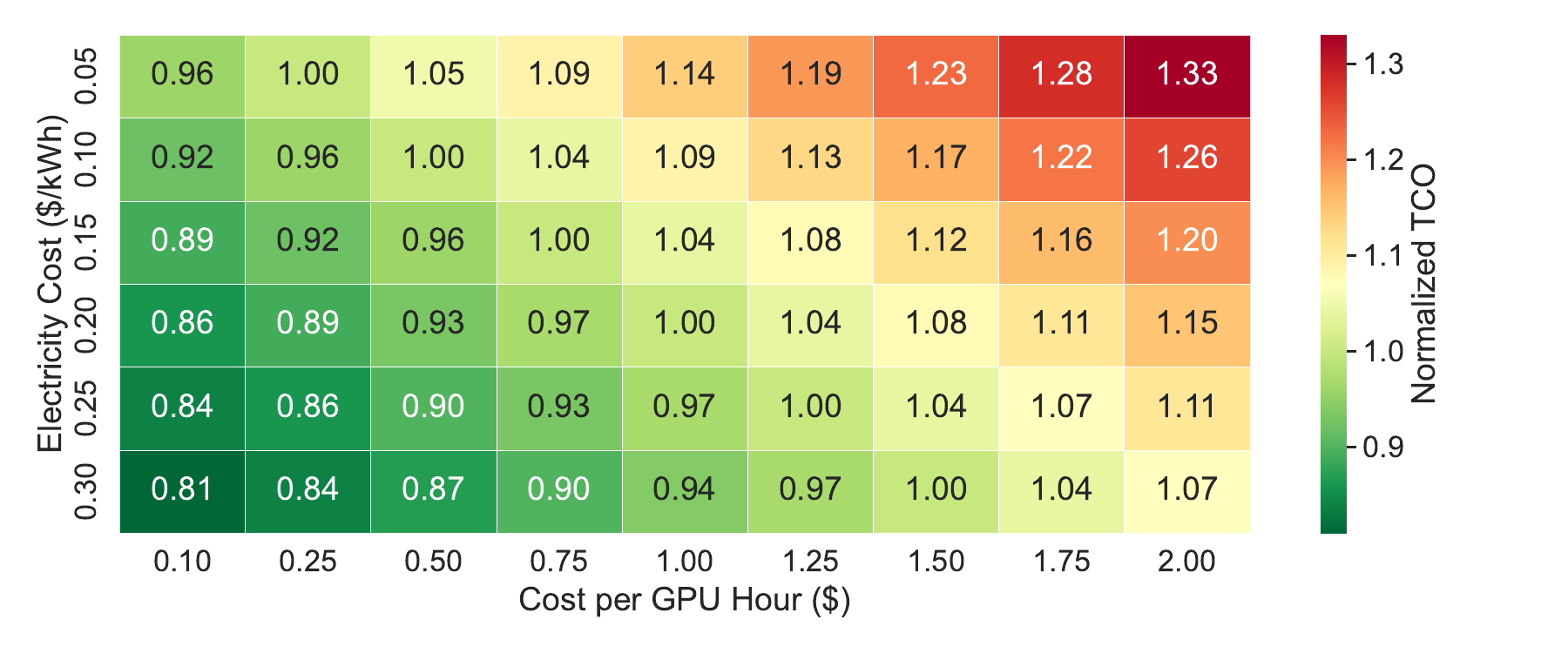}
    \caption{Normalized TCO w.r.t. baseline data center shows a fundamental trade-off of opportunity cost. }
    \label{fig:TCO-heatmap}
    \vspace{-2mm}
    \Description{Total cost of ownership with respect to baseline data center.}
\end{figure}
\section{Related Work} \label{sec:related_Work}

\textbf{Frequency regulation in data centers:}   
Previous data center frequency regulation service works target techniques such as DVFS, stored energy devices, and coordinating local power generators, primarily target batch workloads and operate on 15-minute intervals, which are insufficient for the 2-second adjustments required for frequency regulation~\cite{chen-energyqare:-2019,chen-data-2014,Chen2013,Chen2014,jahanshahi2022powermorph,wang2019frequency,183116}. PowerMorph~\cite{jahanshahi2022powermorph} addresses this with millisecond-level QoS. However, it targets only CPUs, which limits the amount of regulation provision in modern data centers. Additionally, using UPS systems for regulation services is costly and reduces their lifespan, as they are primarily designed for backup power~\cite{6307771}. 

\textbf{Other demand response in data centers: } 
Significant prior works have explored data center demand response services at hourly-granularity to improve the reliability of power grids under stress, such as load shedding during power emergencies or load shifting. There are several strategies, such as temporal and spatial load shifting/job migration~\cite{liu2011,xing2023carbonresponder,dou2017carbon,luo2013temporal,limitations,google_demand_response}, voluntary load reduction and power capping ~\cite{Adaptive-DVFS}, through techniques like Dynamic Voltage and Frequency Scaling (DVFS)~\cite{wu-precise-2018,Cochran2011-PackAndCap}, thread packing~\cite{Cochran2011-PackAndCap, chow2021energy}, and co-scheduling~\cite{hsu-smoothoperator-2018, jahanshahi2020gpu, kiran2}. These traditional data center demand response mechanisms operates at an hour-scale, making it too slow for second-scale frequency regulation.

\textbf{Carbon-aware data centers: }
Many prior studies, have explored carbon-aware policies, such as temporal and spatial (geographical) workload shifting, carbon-aware scheduling, net-zero data centers, etc.~\cite{Li2012-ISwitch,goiri2011greenslot,Li2011-SolarCore,Li2013-Chameleon,liu2011,dou2017carbon, luo2013temporal,lin2023adapting,CarbonScaler,10.1145/3698038.3698542, limitations}. More recently, Carbon Explorer \cite{acun2023carbon-explorer} explores the trade-off between reducing operational carbon and increasing embodied carbon from infrastructure manufacturing. Carbon Responder~\cite{xing2023carbonresponder}, proposed a framework that develops performance-aware power allocation policies to reduce carbon emissions during periods of high grid carbon intensity. While these prior works are aware of the power grid's carbon intensity, these prior works do not actively participate to improve grid reliability or account for carbon emissions of regulation reserves.

\section{Conclusion}\label{sec:conclusion}
This work demonstrates how data centers can decrease reliance on fossil fuel-based regulation reserves. We introduce Exogenous Carbon to capture the grid-side carbon savings due to data center regulation service.
We introduced EcoCenter, a framework to maximize regulation provision and provide accurate regulation service. 
EcoCenter can achieve high-quality regulation, lower TCO, and in certain scenarios, achieve grid-side exogenous carbon savings that completely outweighs the operational carbon of the data center.

\begin{acks}
We would like to thank the anonymous reviewers for their invaluable comments and suggestions. This work is partly supported by the University of California, Riverside, the National Science Foundation under grants CNS-1955650, CNS-2047521, CCF-2324940, and CCF-2324941.
\end{acks}

\newpage
\bibliographystyle{ACM-Reference-Format}
\bibliography{references}

@misc{FreqRegNews,
  author = "GreenTechMedia",
  title = "In California, Solar and Wind Boost the Price of Frequency Regulation",
  url = "http://www.greentechmedia.com/articles/read/in-california-solar-and-wind-boosts-the-price-for-frequency-regulation",
  year = {2022}
}

@article{radford2019language,
  title={Language models are unsupervised multitask learners},
  author={Radford, Alec and Wu, Jeffrey and Child, Rewon and Luan, David and Amodei, Dario and Sutskever, Ilya and others},
  journal={OpenAI blog},
  volume={1},
  number={8},
  pages={9},
  year={2019}
}

@misc{google_report,
  author = "Google",
  title = "Environmental Report",
  url = "https://www.gstatic.com/gumdrop/sustainability/google-2023-environmental-report.pdf",
  year = {2023}
}

@misc{ms_report,
  author = "Microsoft",
  title = "Powering sustainable transformation",
  url = "https://datacenters.microsoft.com/globe/powering-sustainable-transformation/",
  year = {2025}
}

@misc{google_demand_response,
  author = "Google",
  title = "Supporting power grids with demand response at Google data centers",
  url = "https://cloud.google.com/blog/products/infrastructure/using-demand-response-to-reduce-data-center-power-consumption",
  year = {2023}
}

@INPROCEEDINGS{7039172,

  author={Wierman, Adam and Liu, Zhenhua and Liu, Iris and Mohsenian-Rad, Hamed},

  booktitle={International Green Computing Conference}, 

  title={Opportunities and challenges for data center demand response}, 

  year={2014},

  volume={},

  number={},

  pages={1-10},

  doi={10.1109/IGCC.2014.7039172}}

@misc{morgan,
  author = "Morgan Stanley",
  title ="" ,
  url ="https://www.datacenterdynamics.com/en/news/morgan-stanley-data-center-industry-will-emit-25bn-tons-of-co2-by-2030/",
  year ={2023}
}

@misc{renewable_data,
	title = {Electricity Data Browser},
	howpublished = {\url{{https://www.eia.gov/electricity/data/browser/}}},
	urldate = {2023},
	author = {U.S. Energy Information Administration (EIA) },
	year = {2023}
}

@inproceedings{jahanshahi2022scaleserve,
  title={Scaleserve: A scalable multi-gpu machine learning inference system and benchmarking suite},
  author={Jahanshahi, Ali and Chow, Marcus and Wong, Daniel},
  booktitle={Proceedings of the 14th Workshop on General Purpose Processing Using GPU},
  pages={1--2},
  year={2022}
}

@inproceedings {Elfen,
author = {Xi Yang and Stephen M. Blackburn and Kathryn S. McKinley},
title = {Elfen Scheduling: Fine-Grain Principled Borrowing from Latency-Critical Workloads Using Simultaneous Multithreading},
booktitle = {2016 {USENIX} Annual Technical Conference ({USENIX} {ATC} 16)},
year = {2016},
month = jun,
}

@misc{swimprojectucb_2013,
title={SWIM Project},
url={https://github.com/SWIMProjectUCB/SWIM/wiki},
journal={GitHub},
author={SWIMProjectUCB},
year={2013}
}

@INPROCEEDINGS{Chou2019-uDPM,
  title     = "{$\mu$DPM}: Dynamic Power Management for the Microsecond Era",
  booktitle = "2019 {IEEE} International Symposium on High Performance Computer
               Architecture ({HPCA})",
  author    = "Chou, C and Bhuyan, L N and Wong, D",
  pages     = "120--132",
  month     =  feb,
  year      =  2019
}

@inproceedings{de2008z3,
  title={Z3: An efficient SMT solver},
  author={De Moura, Leonardo and Bj{\o}rner, Nikolaj},
  booktitle={International conference on Tools and Algorithms for the Construction and Analysis of Systems},
  pages={337--340},
  year={2008},
  organization={Springer}
}

@misc{rocm-smi,
	title = {Rocm system management interface (rocm smi) library},
	howpublished = {\url{{https://github.com/ROCm/ROC-smi}}},
	urldate = {2024},
	author = {AMD},
	year = {2024}
}

@article{jahanshahi2022powermorph,
  title={PowerMorph: QoS-aware server power reshaping for data center regulation service},
  author={Jahanshahi, Ali and Yu, Nanpeng and Wong, Daniel},
  journal={ACM Transactions on Architecture and Code Optimization (TACO)},
  volume={19},
  number={3},
  pages={1--27},
  year={2022},
  publisher={ACM New York, NY}
}

@article{jahanshahi2020gpu,
  title={Gpu-nest: Characterizing energy efficiency of multi-gpu inference servers},
  author={Jahanshahi, Ali and Sabzi, Hadi Zamani and Lau, Chester and Wong, Daniel},
  journal={IEEE Computer Architecture Letters},
  volume={19},
  number={2},
  pages={139--142},
  year={2020},
  publisher={IEEE}
}

@inproceedings{li2023toward,
  title={Toward sustainable hpc: Carbon footprint estimation and environmental implications of hpc systems},
  author={Li, Baolin and Basu Roy, Rohan and Wang, Daniel and Samsi, Siddharth and Gadepally, Vijay and Tiwari, Devesh},
  booktitle={Proceedings of the International Conference for High Performance Computing, Networking, Storage and Analysis},
  pages={1--15},
  year={2023}
}

@INPROCEEDINGS{wattwiser23,
title={WattWiser: Power \& Resource-Efficient Scheduling for Multi-Model Multi-GPU Inference Servers},
author={Jahanshahi, Ali and Rezvani, Mohammadreza and Wong, Daniel},
booktitle = {2023 IEEE 14th International Green and Sustainable Computing Conference (IGSC)},
year=2023,
}

@inproceedings{acun2023carbon-explorer,
  title={Carbon explorer: A holistic framework for designing carbon aware datacenters},
  author={Acun, Bilge and Lee, Benjamin and Kazhamiaka, Fiodar and Maeng, Kiwan and Gupta, Udit and Chakkaravarthy, Manoj and Brooks, David and Wu, Carole-Jean},
  booktitle={Proceedings of the 28th ACM International Conference on Architectural Support for Programming Languages and Operating Systems, Volume 2},
  pages={118--132},
  year={2023}
}

@misc{pjm-pjm-2019,
	title = {{PJM} {Manual} 12: {Balancing} {Operations}},
	howpublished = {\url{{https://www.pjm.com/-/media/documents/manuals/m12.ashx}}},
	urldate = {2022},
	author = {PJM},
	year = {2022}
}

@article{dou2017carbon,
  title={Carbon-aware electricity cost minimization for sustainable data centers},
  author={Dou, Hui and Qi, Yong and Wei, Wei and Song, Houbing},
  journal={IEEE Transactions on Sustainable Computing},
  volume={2},
  number={2},
  pages={211--223},
  year={2017},
  publisher={IEEE}
}

@article{luo2013temporal,
  title={Temporal load balancing with service delay guarantees for data center energy cost optimization},
  author={Luo, Jianying and Rao, Lei and Liu, Xue},
  journal={IEEE Transactions on Parallel and Distributed Systems},
  volume={25},
  number={3},
  pages={775--784},
  year={2013},
  publisher={IEEE}
}

@inproceedings{lin2023adapting,
  title={Adapting Datacenter Capacity for Greener Datacenters and Grid},
  author={Lin, Liuzixuan and Chien, Andrew A},
  booktitle={Proceedings of the 14th ACM International Conference on Future Energy Systems},
  pages={200--213},
  year={2023}
}

@article{cook2019clicking,
  title={Clicking Clean Virginia The Dirty Energy Powering Data Center Alley},
  author={Cook, Gary and Jardim, Elizabeth and Craighill, C},
  journal={Greenpeace Available from: https://www. greenpeace. org/usa/reports/click-clean-virginia/[Accessed 21 Mar 2020]},
  year={2019}
}

@inproceedings{li2023clover,
  title={Clover: Toward Sustainable AI with Carbon-Aware Machine Learning Inference Service},
  author={Li, Baolin and Samsi, Siddharth and Gadepally, Vijay and Tiwari, Devesh},
  booktitle={Proceedings of the International Conference for High Performance Computing, Networking, Storage and Analysis},
  pages={1--15},
  year={2023}
}

@article{touvron2023llama,
  title={Llama 2: Open foundation and fine-tuned chat models},
  author={Touvron, Hugo and Martin, Louis and Stone, Kevin and Albert, Peter and Almahairi, Amjad and Babaei, Yasmine and Bashlykov, Nikolay and Batra, Soumya and Bhargava, Prajjwal and Bhosale, Shruti and others},
  journal={arXiv preprint arXiv:2307.09288},
  year={2023}
}

@inproceedings{Adaptive-DVFS,
author = {Cochran, Ryan and Hankendi, Can and Coskun, Ayse K. and Reda, Sherief},
title = {Pack \& Cap: adaptive DVFS and thread packing under power caps},
year = {2011},
booktitle = {Proceedings of the 44th Annual IEEE/ACM International Symposium on Microarchitecture},
series = {MICRO-44}
}

@misc{gorka2024carbonintensity,
      title={ElectricityEmissions.jl: A Framework for the Comparison of Carbon Intensity Signals}, 
      author={Joe Gorka and Noah Rhodes and Line Roald},
      year={2024},
      eprint={2411.06560},
      archivePrefix={arXiv},
      primaryClass={eess.SY},
      url={https://arxiv.org/abs/2411.06560}, 
}

@misc{electricitymaps,
      title={The leading API for granular electricity data Reduce carbon emissions with actionable electricity data.},
      url = {https://www.electricitymaps.com},
year = {2024}
}

@misc{GSprojection,
      title={AI is poised to drive 160\% increase in data center power demand},
      url = {https://www.goldmansachs.com/insights/articles/AI-poised-to-drive-160-increase-in-power-demand},
year = {2024}
}

@inproceedings{Sukprasert2024,
author = {Sukprasert, Thanathorn and Souza, Abel and Bashir, Noman and Irwin, David and Shenoy, Prashant},
title = {On the Limitations of Carbon-Aware Temporal and Spatial Workload Shifting in the Cloud},
year = {2024},
booktitle = {Proceedings of the Nineteenth European Conference on Computer Systems},
series = {EuroSys '24}
}

@article{Anderson2025, 
year = {2025}, 
title = {{Optimizing deep decarbonization pathways in California with power system planning using surrogate level-based Lagrangian relaxation}}, 
author = {Anderson, Osten and Bragin, Mikhail A. and Yu, Nanpeng}, 
journal = {Applied Energy}, 
issn = {0306-2619}, 
volume = {377}, 
}

@inproceedings{Chen2014,
  title = {Reducing the Data Center Electricity Costs through Participation in Smart Grid Programs},


  booktitle = {International {{Green Computing Conference}}},
  author = {Chen, H. and Caramanis, M. C. and Coskun, A. K.},
  month = nov,
  year = {2014},

  pages = {1-10},

}

@inproceedings{Chen2013,
  address = {Piscataway, NJ, USA},
  series = {ICCAD '13},
  title = {Dynamic {{Server Power Capping}} for {{Enabling Data Center Participation}} in {{Power Markets}}},
  isbn = {978-1-4799-1069-4},

  booktitle = {Proceedings of the {{International Conference}} on {{Computer}}-{{Aided Design}}},
  publisher = {{IEEE Press}},
  author = {Chen, Hao and Hankendi, Can and Caramanis, Michael C. and Coskun, Ayse K.},
  year = {2013},
  pages = {122--129},

}

@inproceedings{chen-data-2014,
	address = {Singapore},
	title = {The data center as a grid load stabilizer},
	isbn = {978-1-4799-2816-3},
	url = {http://ieeexplore.ieee.org/document/6742874/},
	doi = {10.1109/ASPDAC.2014.6742874},
	urldate = {2019-01-15},
	booktitle = {19th {Asia} and {South} {Pacific} {Design} {Automation} {Conference} ({ASP}-{DAC})},
	publisher = {IEEE},
	author = {Chen, Hao and Caramanis, Michael C. and Coskun, Ayse K.},
	month = jan,
	year = {2014},
	pages = {105--112},

}

@article{chen-energyqare:-2019,
	title = {{EnergyQARE}: {QoS}-{Aware} {Data} {Center} {Participation} in {Smart} {Grid} {Regulation} {Service} {Reserve} {Provision}},
	volume = {4},
	issn = {2376-3639},
	shorttitle = {{EnergyQARE}},
	url = {http://doi.acm.org/10.1145/3243172},
	doi = {10.1145/3243172},
	number = {1},
	urldate = {2019-03-27},
	journal = {ACM Trans. Model. Perform. Eval. Comput. Syst.},
	author = {Chen, Hao and Zhang, Yijia and Caramanis, Michael C. and Coskun, Ayse K.},
	month = jan,
	year = {2019},
	pages = {2:1--2:31},

}

@inproceedings{zhang2019data,
  title={Data Center Participation in Demand Response Programs with Quality-of-Service Guarantees},
  author={Zhang, Yijia and Paschalidis, Ioannis Ch and Coskun, Ayse K},
  booktitle={Proceedings of the Tenth ACM International Conference on Future Energy Systems},
  pages={285--302},
  year={2019}
}

@ARTICLE{zhang2022hpc,
  author={Zhang, Yijia and Wilson, Daniel Curtis and Paschalidis, Ioannis Ch. and Coskun, Ayse K.},
  journal={IEEE Transactions on Sustainable Computing}, 
  title={HPC Data Center Participation in Demand Response: An Adaptive Policy With QoS Assurance}, 
  year={2022},
  volume={7},
  number={1},
  pages={157-171},
}

@article{wang2019frequency,
  title={Frequency regulation service provision in data center with computational flexibility},
  author={Wang, Wei and Abdolrashidi, Amirali and Yu, Nanpeng and Wong, Daniel},
  journal={Applied Energy},
  volume={251},
  pages={113304},
  year={2019},
  publisher={Elsevier}
}

@INPROCEEDINGS{Li2013-Chameleon,
author={C. {Li} and X. {Li} and R. {Wang} and T. {Li} and N. {Goswami} and D. {Qian}},
booktitle={International Symposium on Low Power Electronics and Design (ISLPED)},
title={Chameleon: Adapting throughput server to time-varying green power budget using online learning},
year={2013},
volume={},
number={},
pages={100-105},
doi={10.1109/ISLPED.2013.6629274},
ISSN={null},
month={Sep.},}

@INPROCEEDINGS{Li2011-SolarCore,
author={C. {Li} and W. {Zhang} and C. {Cho} and T. {Li}},
booktitle={2011 IEEE 17th International Symposium on High Performance Computer Architecture},
title={SolarCore: Solar energy driven multi-core architecture power management},
year={2011},
volume={},
number={},
pages={205-216},
doi={10.1109/HPCA.2011.5749729},
ISSN={1530-0897},
month={Feb},}

@inproceedings{goiri2011greenslot,
  title={Greenslot: scheduling energy consumption in green datacenters},
  author={Goiri, {\'I}{\~n}igo and Le, Kien and Haque, Md E and Beauchea, Ryan and Nguyen, Thu D and Guitart, Jordi and Torres, Jordi and Bianchini, Ricardo},
  booktitle={International Conference for High Performance Computing, Networking, Storage and Analysis},
  year={2011}
}

@inproceedings{Li2012-ISwitch,
author = {Li, Chao and Qouneh, Amer and Li, Tao},
title = {ISwitch: Coordinating and Optimizing Renewable Energy Powered Server Clusters},
year = {2012},
isbn = {9781450316422},
publisher = {IEEE Computer Society},
address = {USA},
booktitle = {Proceedings of the 39th Annual International Symposium on Computer Architecture},
pages = {512–523},
numpages = {12},
location = {Portland, Oregon},
series = {ISCA ’12}
}

@inproceedings {183116,
author = {Chao Li and Rui Wang and Tao Li and Depei Qian and Jingling Yuan},
title = {Managing Green Datacenters Powered by Hybrid Renewable Energy Systems},
booktitle = {11th International Conference on Autonomic Computing ({ICAC} 14)},
year = {2014},
isbn = {978-1-931971-11-9},
pages = {261--272},
month = jun,
}

@INPROCEEDINGS{6307771,
author={S. {Govindan} and A. {Sivasubramaniam} and B. {Urgaonkar}},
booktitle={2011 38th Annual International Symposium on Computer Architecture (ISCA)},
title={Benefits and limitations of tapping into stored energy for datacenters},
year={2011},
volume={},
number={},
pages={341-351},
doi={},
ISSN={1063-6897},
month={June},}

@article{aljbour2024powering,
  title={Powering Intelligence: Analyzing Artificial Intelligence and Data Center Energy Consumption},
  author={Aljbour, Jordan and Wilson, Tom and Patel, P},
  journal={EPRI White Paper},
  year={2024}
}

@misc{xing2023carbonresponder,
      title={Carbon Responder: Coordinating Demand Response for the Datacenter Fleet}, 
      author={Jiali Xing and Bilge Acun and Aditya Sundarrajan and David Brooks and Manoj Chakkaravarthy and Nikky Avila and Carole-Jean Wu and Benjamin C. Lee},
      year={2023},
      eprint={2311.08589},
      archivePrefix={arXiv},
      primaryClass={cs.DC},
      url={https://arxiv.org/abs/2311.08589}, 
}

@article{liu2011,
author = {Liu, Zhenhua and Lin, Minghong and Wierman, Adam and Low, Steven H. and Andrew, Lachlan L.H.},
title = {Geographical load balancing with renewables},
year = {2011},
issue_date = {December 2011},
publisher = {Association for Computing Machinery},
address = {New York, NY, USA},
volume = {39},
number = {3},
issn = {0163-5999},
url = {https://doi.org/10.1145/2160803.2160862},
doi = {10.1145/2160803.2160862},
journal = {SIGMETRICS Perform. Eval. Rev.},
month = dec,
pages = {62–66},
numpages = {5}
}

@article{wu-precise-2018,
	title = {Precise {Power} {Capping} for {Latency}-{Sensitive} {Applications} in {Datacenter}},
	issn = {2377-3782},
	doi = {10.1109/TSUSC.2018.2881893},
	journal = {IEEE Transactions on Sustainable Computing},
	author = {Wu, S. and Chen, Y. and Wang, X. and Jin, H. and Liu, F. and Chen, H. and Yan, C.},
	year = {2018},
	pages = {1--1}
}

@inproceedings{hsu-smoothoperator-2018,
	title = {{SmoothOperator}: {Reducing} {Power} {Fragmentation} and {Improving} {Power} {Utilization} in {Large}-scale {Datacenters}},
	booktitle = {23rd {International} {Conference} on {Architectural} {Support} for {Programming} {Languages} and {Operating} {Systems}},
	author = {Hsu, Chang-Hong and Deng, Qingyuan and Mars, Jason and Tang, Lingjia},
	year = {2018}
}

@INPROCEEDINGS{Cochran2011-PackAndCap,
  title     = "Pack \& Cap: Adaptive {DVFS} and Thread Packing Under Power Caps",
  booktitle = "44th Annual {IEEE/ACM} International
               Symposium on Microarchitecture",
  author    = "Cochran, Ryan and Hankendi, Can and Coskun, Ayse K and Reda,
               Sherief",
  year      =  2011
}

@inproceedings{chow2021energy,
  title={Energy Efficient Task Graph Execution Using Compute Unit Masking in GPUs},
  author={Chow, Marcus and Ranganath, Kiran and Lerias, Robert and Carodan, Mika Shanela and Wong, Daniel},
  booktitle={Redefining Scalability for Diversely Heterogeneous Architectures Workshop (RSDHA)},
  year={2021}
}

@inproceedings{kiran2,
  title={Mapa: Multi-accelerator pattern allocation policy for multi-tenant gpu servers},
  author={Ranganath, Kiran and Suetterlein, Joshua D and Manzano, Joseph B and Song, Shuaiwen Leon and Wong, Daniel},
  booktitle={Proceedings of the International Conference for High Performance Computing, Networking, Storage and Analysis},
  pages={1--14},
  year={2021}
}

@article{CarbonScaler,
author = {Hanafy, Walid A. and Liang, Qianlin and Bashir, Noman and Irwin, David and Shenoy, Prashant},
title = {CarbonScaler: Leveraging Cloud Workload Elasticity for Optimizing Carbon-Efficiency},
year = {2023},
issue_date = {December 2023},
publisher = {Association for Computing Machinery},
address = {New York, NY, USA},
volume = {7},
number = {3},
url = {https://doi.org/10.1145/3626788},
doi = {10.1145/3626788},
journal = {Proc. ACM Meas. Anal. Comput. Syst.},
month = dec,
articleno = {57},
numpages = {28}
}

@inproceedings{10.1145/3698038.3698542,
author = {Bashir, Noman and Gohil, Varun and Subramanya, Anagha Belavadi and Shahrad, Mohammad and Irwin, David and Olivetti, Elsa and Delimitrou, Christina},
title = {The Sunk Carbon Fallacy: Rethinking Carbon Footprint Metrics for Effective Carbon-Aware Scheduling},
year = {2024},
isbn = {9798400712869},
publisher = {Association for Computing Machinery},
address = {New York, NY, USA},
url = {https://doi.org/10.1145/3698038.3698542},
doi = {10.1145/3698038.3698542},
booktitle = {Proceedings of the 2024 ACM Symposium on Cloud Computing},
pages = {542–551},
numpages = {10},
location = {Redmond, WA, USA},
series = {SoCC '24}
}

@inproceedings{limitations,
author = {Sukprasert, Thanathorn and Souza, Abel and Bashir, Noman and Irwin, David and Shenoy, Prashant},
title = {On the Limitations of Carbon-Aware Temporal and Spatial Workload Shifting in the Cloud},
year = {2024},
isbn = {9798400704376},
publisher = {Association for Computing Machinery},
address = {New York, NY, USA},
url = {https://doi.org/10.1145/3627703.3650079},
doi = {10.1145/3627703.3650079},
booktitle = {Proceedings of the Nineteenth European Conference on Computer Systems},
pages = {924–941},
numpages = {18},
location = {Athens, Greece},
series = {EuroSys '24}
}

@inproceedings{LLM-Power-Management,
author = {Patel, Pratyush and Choukse, Esha and Zhang, Chaojie and Goiri, \'{I}\~{n}igo and Warrier, Brijesh and Mahalingam, Nithish and Bianchini, Ricardo},
title = {Characterizing Power Management Opportunities for LLMs in the Cloud},
year = {2024},
isbn = {9798400703867},
publisher = {Association for Computing Machinery},
address = {New York, NY, USA},
url = {https://doi.org/10.1145/3620666.3651329},
doi = {10.1145/3620666.3651329},
booktitle = {Proceedings of the 29th ACM International Conference on Architectural Support for Programming Languages and Operating Systems, Volume 3},
pages = {207–222},
numpages = {16},
location = {La Jolla, CA, USA},
series = {ASPLOS '24}
}

@article{swierczynski_field_2014,
	title = {Field {Experience} from {Li}-{Ion} {BESS} {Delivering} {Primary} {Frequency} {Regulation} in the {Danish} {Energy} {Market}},
	volume = {61},
	doi = {10.1149/06137.0001ecst},
	number = {37},
	journal = {ECS Transactions},
	author = {Świerczyński, Maciej and Stroe, Daniel Ioan and Lærke, Rasmus and Stan, Ana Irina and Kjær, Philip Carne and Teodorescu, Remus and Kær, Soren Knudsen},
	month = sep,
	year = {2014},
}

@misc{A100,
	title = {NVIDIA A100 Tensor Core GPU
Architecture},
	howpublished = {\url{{https://images.nvidia.com/aem-dam/en-zz/Solutions/data-center/nvidia-ampere-architecture-whitepaper.pdf}}},
	urldate = {2020},
	author = {NVIDIA},
	year = {2020}
}

@misc{H100,
	title = {NVIDIA H100 Tensor Core GPU
Architecture},
	howpublished = {\url{{https://www.advancedclustering.com/wp-content/uploads/2022/03/gtc22-whitepaper-hopper.pdf}}},
	urldate = {2022},
	author = {NVIDIA},
	year = {2022}
}

@misc{B200,
	title = {NVIDIA Blackwell Architecture
Technical Brief},
	howpublished = {\url{{https://cdn.prod.website-files.com/61dda201f29b7efc52c5fbaf/6602ea9d0ce8cb73fb6de87f\_nvidia-blackwell-architecture-technical-brief.pdf}}},
	urldate = {2024},
	author = {NVIDIA},
	year = {2024}
}

@book{41606,title	= {The Datacenter as a Computer: An Introduction to the Design of Warehouse-Scale Machines, Second Edition},author	= {Luiz André Barroso and Jimmy Clidaras and Urs Hölzle},year	= {2013}, publisher={Morgan \& Claypool Publishers}
}

@misc{google_PUE,
  author = "Google",
  title = "Google data center PUE performance ",
  url = "https://datacenters.google/efficiency/",
  year = {2025}
}

@misc{exxact,
  author = "Exxact Corporation",
  title = "AMD Instinct MI300 Series Systems",
  url = "https://www.exxactcorp.com/category/AMD-Radeon-Instinct-Solutions",
  year = {2025}
}

@misc{hamilton,
  author = "James Hamilton",
  title = "Overall Data Center Costs",
  url = "https://perspectives.mvdirona.com/2010/09/overall-data-center-costs/",
  year = {2025}
}

@misc{caiso-as,
  author = "CAISO",
  title = "Market Performance Report, April 2024, IFM Procurement and Prices",
  url = "https://www.caiso.com/content/monthly-market-performance/apr-2024/ancillary-services.html",
  year = {2025}
}

@inproceedings{weng2023beware,
  title={Beware of fragmentation: Scheduling $\{$GPU-Sharing$\}$ workloads with fragmentation gradient descent},
  author={Weng, Qizhen and Yang, Lingyun and Yu, Yinghao and Wang, Wei and Tang, Xiaochuan and Yang, Guodong and Zhang, Liping},
  booktitle={2023 USENIX Annual Technical Conference (USENIX ATC 23)},
  pages={995--1008},
  year={2023}
}

@ARTICLE{6881647,
  author={Calheiros, Rodrigo N. and Masoumi, Enayat and Ranjan, Rajiv and Buyya, Rajkumar},
  journal={IEEE Transactions on Cloud Computing}, 
  title={Workload Prediction Using ARIMA Model and Its Impact on Cloud Applications’ QoS}, 
  year={2015},
  volume={3},
  number={4},
  pages={449-458},
  doi={10.1109/TCC.2014.2350475}}

@article{cetinski2015ame,
  title={AME-WPC: Advanced model for efficient workload prediction in the cloud},
  author={Cetinski, Katja and Juric, Matjaz B},
  journal={Journal of Network and Computer Applications},
  volume={55},
  pages={191--201},
  year={2015},
  publisher={Elsevier}
}

@inproceedings{di2012host,
  title={Host load prediction in a Google compute cloud with a Bayesian model},
  author={Di, Sheng and Kondo, Derrick and Cirne, Walfredo},
  booktitle={SC'12: Proceedings of the International Conference on High Performance Computing, Networking, Storage and Analysis},
  pages={1--11},
  year={2012},
  organization={IEEE}
}

@INPROCEEDINGS{6274197,
  author={Fang, Wei and Lu, ZhiHui and Wu, Jie and Cao, ZhenYin},
  booktitle={2012 IEEE Ninth International Conference on Services Computing}, 
  title={RPPS: A Novel Resource Prediction and Provisioning Scheme in Cloud Data Center}, 
  year={2012},
  volume={},
  number={},
  pages={609-616},
  doi={10.1109/SCC.2012.47}}

@article{liu2017adaptive,
  title={An adaptive prediction approach based on workload pattern discrimination in the cloud},
  author={Liu, Chunhong and Liu, Chuanchang and Shang, Yanlei and Chen, Shiping and Cheng, Bo and Chen, Junliang},
  journal={Journal of Network and Computer Applications},
  volume={80},
  pages={35--44},
  year={2017},
  publisher={Elsevier}
}

@INPROCEEDINGS{8102182,
  author={Wamba, Gilles Madi and Li, Yunbo and Orgerie, Anne-Cécile and Beldiceanu, Nicolas and Menaud, Jean-Marc},
  booktitle={2017 29th International Symposium on Computer Architecture and High Performance Computing (SBAC-PAD)}, 
  title={Cloud Workload Prediction and Generation Models}, 
  year={2017},
  volume={},
  number={},
  pages={89-96},
  doi={10.1109/SBAC-PAD.2017.19}}

@ARTICLE{8301555,
  author={Zhang, Qingchen and Yang, Laurence T. and Yan, Zheng and Chen, Zhikui and Li, Peng},
  journal={IEEE Transactions on Industrial Informatics}, 
  title={An Efficient Deep Learning Model to Predict Cloud Workload for Industry Informatics}, 
  year={2018},
  volume={14},
  number={7},
  pages={3170-3178},
  doi={10.1109/TII.2018.2808910}}

@techreport{caiso-ghg,
    author = {Hundiwale, Abhishek},
    institution = {{California ISO}},
    title = {Greenhouse Gas Emission Tracking Methodology},
    year = {2016},
}

@techreport{caiso-ghg2,
    institution = {{California ISO}},
    author = {{California ISO}},
    title = {Greenhouse Gas Coordination Stakeholder Recommendations for Policy Development},
    year = {2024},
}

@ARTICLE{8318697,
  author={Sadeghi-Mobarakeh, Ashkan and Mohsenian-Rad, Hamed},
  journal={IEEE Transactions on Power Systems}, 
  title={Performance Accuracy Scores in CAISO and MISO Regulation Markets: A Comparison Based on Real Data and Mathematical Analysis}, 
  year={2018},
  volume={33},
  number={3},
  pages={3196-3198},
  doi={10.1109/TPWRS.2018.2816808}}

@misc{lacoste2019quantifyingcarbonemissionsmachine,
      title={Quantifying the Carbon Emissions of Machine Learning}, 
      author={Alexandre Lacoste and Alexandra Luccioni and Victor Schmidt and Thomas Dandres},
      year={2019},
      eprint={1910.09700},
      archivePrefix={arXiv},
      primaryClass={cs.CY},
      url={https://arxiv.org/abs/1910.09700}, 
}

@misc{power-min,
  author = "Steven Seachman",
  title = "Process Control Strategies for Reducing the Minimum Load of Fossil-Fired Plants",
  url = "https://www.powermag.com/process-control-strategies-for-reducing-the-minimum-load-of-fossil-fired-plants/",
  year = {2025}
}

@ARTICLE{power-min2,
  author={Yu, Nan-Peng and Liu, Chen-Ching and Price, James},
  journal={IEEE Transactions on Power Systems}, 
  title={Evaluation of Market Rules Using a Multi-Agent System Method}, 
  year={2010},
  volume={25},
  number={1},
  pages={470-479},
}

@article{power-min3,
title = {Feasibility study of Combined Cycle Gas Turbine (CCGT) power plant integration with Adiabatic Compressed Air Energy Storage (ACAES)},
journal = {Applied Energy},
volume = {221},
pages = {477-489},
year = {2018},
author = {Jacek D. Wojcik and Jihong Wang},
}

@techreport{brooks2000ge,
  title={GE gas turbine performance characteristics},
  author={Brooks, Frank J},
  year={2000},
  institution={GE Power Systems}
}

@techreport{LBNL-DR,
  title={2025 California Demand Response Potential Study-Charting California’s Demand Response Future. Final Report on Phase 2 Results},
  author={Alstone, Peter and Potter, Jennifer and Piette, Mary Ann and Schwartz, Peter and Berger, Michael A and Dunn, Laurel N and Smith, Sarah J and Sohn, Michael D and Aghajanzadeh, Aruab and Stensson, Sofia and others},
  year={2017},
  institution={Lawrence Berkeley National Lab.(LBNL), Berkeley, CA (United States)}
}

@inproceedings{lindberg2021guidereducingcarbonemissions,
author = {Lindberg, Julia and Abdennadher, Yasmine and Chen, Jiaqi and Lesieutre, Bernard C. and Roald, Line},
title = {A Guide to Reducing Carbon Emissions through Data Center Geographical Load Shifting},
year = {2021},
doi = {10.1145/3447555.3466582},
booktitle = {Proceedings of the Twelfth ACM International Conference on Future Energy Systems},
series = {e-Energy '21}
}

@misc{nvidia_green_contexts,
  author       = {NVIDIA},
  title        = {Green Contexts - CUDA Driver API},
  year         = {2025},
  url          = {https://docs.nvidia.com/cuda/cuda-driver-api/group__CUDA__GREEN__CONTEXTS.html},
  note         = {Accessed: 2025-10-29}
}

@InProceedings{libsmctrl,
  author =	{Bakita, Joshua and Anderson, James H.},
  title =	{Hardware Compute Partitioning on NVIDIA GPUs for Composable Systems},
  booktitle =	{37th Euromicro Conference on Real-Time Systems (ECRTS 2025)},
  year =	{2025},
  URL =		{https://drops.dagstuhl.de/entities/document/10.4230/LIPIcs.ECRTS.2025.21},
}

@ARTICLE{gpupower-cal,
  author={Patel, Pratyush and Gong, Zibo and Rizvi, Syeda and Choukse, Esha and Misra, Pulkit and Anderson, Thomas and Sriraman, Akshitha},
  journal={IEEE Computer Architecture Letters}, 
  title={Towards Improved Power Management in Cloud GPUs}, 
  year={2023},
  volume={22},
  number={2},
  pages={141-144},
  doi={10.1109/LCA.2023.3278652}}

\end{document}